\documentclass[preprint,showpacs,preprintnumbers]{revtex4-1}
\usepackage{graphicx,latexsym,amssymb,amsmath,color,multirow,mathrsfs,booktabs,ifpdf}
\usepackage{dcolumn}
\usepackage{bm}
\usepackage{longtable}
\def\nA{nucleon-nucleus\ }
\def\nPb{$n+^{208}$Pb\ }

\begin{document}
\title{Extended Hartree-Fock study of the single-particle potential: the nuclear 
symmetry energy, nucleon effective mass, and folding model of the nucleon optical 
potential} 
\author{Doan Thi Loan$^1$}
\author{Bui Minh Loc$^{1,2}$}
\author{Dao T. Khoa$^1$}\email{khoa@vinatom.gov.vn}
\affiliation{$^1$Institute for Nuclear Science and Technique, VINATOM \\ 
179 Hoang Quoc Viet, Cau Giay, Hanoi, Vietnam.\\ 
$^2$ University of Pedagogy, Ho Chi Minh City, Vietnam.}
\begin{abstract}
{The nucleon mean-field potential has been thoroughly investigated in an 
extended Hartree-Fock (HF) calculation of nuclear matter (NM) using the CDM3Y3 
and CDM3Y6 density dependent versions of the M3Y interaction. The single-particle 
(s/p) energies of nucleons in NM are determined according to the Hugenholtz-van 
Hove theorem, which gives rise naturally to a rearrangement term (RT) of the s/p 
potential at the Fermi momentum. Using the RT obtained exactly at the different NM 
densities and neutron-proton asymmetries, a consistent method is suggested to 
take into account effectively the momentum dependence of the RT of the s/p 
potential within the standard HF scheme. To obtain a realistic 
momentum dependence of the nucleon optical potential (OP), the high-momentum 
part of the s/p potential was accurately readjusted to reproduce 
the observed energy dependence of the nucleon OP over a wide range of energies. 
The impact of the RT and momentum dependence of the s/p potential on the 
density dependence of the nuclear symmetry energy and nucleon effective mass 
has been studied in details. The high-momentum tail of the s/p potential was 
found to have a sizable effect on the slope of the symmetry energy and the 
neutron-proton effective mass splitting at supranuclear densities of the NM. 
Based on a local density approximation, the folding model of the nucleon OP 
of finite nuclei has been extended to take into account consistently the 
RT and momentum dependence of the nucleon OP in the same mean-field manner, 
and successfully applied to study the elastic neutron scattering on the 
lead target at the energies around the Fermi energy.}
\end{abstract}
\date{accepted for publication in Phys. Rev. C, August 21, 2015}
\maketitle

\section{Introduction}
\label{intro} 
The nucleon mean-field potential or single-particle (s/p) potential is the most 
important quantity determining the s/p properties of neutrons and protons in the 
nuclear matter (NM) as well as in the finite nuclei \cite{Mah85},
and it has been on the focus of recent many-body studies of the NM, like the 
Brueckner-Hartree-Fock (BHF) calculations of the NM starting from a realistic 
choice of the free nucleon-nucleon (NN) interaction \cite{Bom91,Zuo99,Zuo14} 
or the mean-field studies of the NM on the Hartree-Fock (HF) level, using the
different versions of the effective (in-medium) NN interaction \cite{Xu10,Che12,Xu14}. 
Quite interesting aspect of the s/p potential is its direct connection with the 
nuclear symmetry energy \cite{Xu14,Zuo14}, a key quantity necessary for the 
determination of the equation of state (EOS) of the asymmetric NM \cite{Hor14}. 
Many microscopic studies of the EOS were done based either on
the nonrelativistic or relativistic mean-field potential given by realistic 
two-body and three-body NN forces or interaction Lagrangians \cite{Ba08,Bal07}. 
Such microscopic many-body studies did show the important role played by the
Pauli blocking effects as well as the increasing strength of the higher-order NN 
correlations at the high NM densities. These medium effects are usually considered 
as the main physics origin of an explicit density dependence embedded 
in the different versions of the effective NN interaction, being used currently
in the HF calculations of the nuclear structure or nuclear reaction studies. 
Among them, quite popular are the density dependent versions of the M3Y 
interaction (originally constructed to reproduce the G-matrix elements 
of the Reid \cite{Be77} and Paris \cite{An83} NN potentials in an oscillator 
basis), which have been successfully used in the HF studies of the NM 
\cite{Kho93,Kho95,Kho96,Tha09,Loa11} as well as in the folding model 
studies of the nucleon-nucleus and nucleus-nucleus scattering 
\cite{Kho97,Kho02,Kho07,Kho07r,Kho09,Kho14}.

With the phenomenological density dependence of the M3Y interaction parametrized 
to give a realistic description of the NM saturation properties 
\cite{Kho93,Kho95,Kho97} within the HF frame, the HF nucleon optical potential 
(or the high-momentum part of the s/p potential) was used to determine the explicit
energy dependence of the density dependent M3Y interaction \cite{Kho93,Kho95s} 
based on the observed energy dependence of the nucleon optical potential (OP). 
These density- and energy dependent versions of the M3Y interaction have been further 
used in the folding model calculation of the nucleon-nucleus and nucleus-nucleus 
OP \cite{Kho02,Kho07,Kho07r,Kho09}. The simple assumption for the s/p potential in 
the NM made in Refs.~\cite{Kho93,Kho95s} is roughly equivalent to the microscopic 
s/p potential of the Brueckner-Bethe theory \cite{Brown}, which lacks the so-called 
\emph{rearrangement} term that arises naturally in the Landau theory for the infinite 
Fermi systems \cite{Migdal}. Such a rearrangement term (RT) also appears when the s/p 
potential is evaluated from the total NM energy using the Hugenholtz and van Hove 
(HvH) theorem \cite{HvH}, which is exact for all the interacting Fermi systems, 
independent of the type of the interaction between fermions. 
For the infinite NM, it is straightforward to see that the HvH theorem is satisfied 
on the HF level only when the in-medium NN interaction is density independent, i.e., 
when the RT is equal zero \cite{Cze02}. It is, therefore, of high interest to assess 
the impact of the RT on the s/p potential in a mean-field study of the NM within 
the standard HF frame using a realistic density dependent NN interaction. Moreover,
given the fact that the nuclear symmetry energy and nucleon effective mass are 
directly linked to the momentum- and density dependence of the single-nucleon 
potential \cite{Xu14,Zuo14}, it is highly desirable to have a method to take 
into account properly the density- and momentum dependence of the RT of the s/p 
potential on the HF level, which is the main motivation of the present study. It is
further expected that, in a local density approximation, one should be able to 
include, in the same mean-field manner, the RT and momentum dependence 
of the density dependent NN interaction into the folding model calculation of the 
nucleon OP for the finite nuclei, which is being evaluated so far mostly on the HF level 
\cite{Kho02,Kho07,Kho09,Kho14}. Towards this goal, an extension of the single-folding
approach for the nucleon OP \cite{Kho02} has been done and applied to study the elastic 
neutron scattering on the lead target measured at the incident energies of 30.4 and 40 MeV
\cite{Dev12}. 

\section{Extended HF formalism for the single-particle potential}
\label{sec1}
In the present HF approach, we consider the homogeneous, spin-saturated NM 
of given neutron ($\rho_n$) and proton ($\rho_p$) densities, or equivalently, 
of given total nucleon density $\rho=\rho_n+\rho_p$ and neutron-proton (n/p) 
asymmetry $\delta=(\rho_n-\rho_p)/\rho$. Using the direct ($v^{\rm D}_{\rm c}$) 
and exchange ($v^{\rm EX}_{\rm c}$) parts of the (central) in-medium NN interaction 
$v_{\rm c}$, the total NM energy can be determined in the standard nonrelativistic 
HF scheme as $E=E_{\rm kin}+E_{\rm pot}$, where the kinetic and potential energies are
\begin{eqnarray}
 E_{\rm kin} &=& \sum_{k \sigma \tau} n_\tau(k)
 \frac{\hbar^{2}k^2}{2m_\tau}  \label{ek1} \\
 E_{\rm pot} &=& {\frac{1}{ 2}}\sum_{k \sigma \tau} \sum_{k'\sigma '\tau '}
 n_{\tau}(k) n_{\tau'}(k')
[\langle{\bm{k}\sigma\tau,\bm{k}'\sigma'\tau'}|v^{\rm D}_{\rm c}|
{\bm{k}\sigma\tau,\bm{k}'\sigma'\tau'}\rangle \nonumber \\
& & \hskip 2cm +\ \langle{\bm{k}\sigma\tau,\bm{k}'\sigma'\tau'}
|v^{\rm EX}_{\rm c}|{\bm{k}'\sigma\tau,\bm{k}\sigma'\tau'}\rangle] \nonumber \\
&=& {\frac{1}{ 2}}\sum_{k \sigma \tau} \sum_{k'\sigma '\tau '}
 n_{\tau}(k)n_{\tau'}(k')
\langle{\bm{k}\sigma\tau,\bm{k}'\sigma'\tau'}|v_{\rm c}|
{\bm{k}\sigma\tau,\bm{k}'\sigma'\tau'}\rangle_\mathcal{A}. \label{ek2} 
\end{eqnarray}
The s/p wave function $|\bm{k}\sigma\tau\rangle$ is plane wave, 
and the summation in (\ref{ek1})-(\ref{ek2}) is done separately over the 
neutron ($\tau=n$) and proton ($\tau=p$) s/p indices. The nucleon 
momentum distribution $n_\tau(k)$ in the spin-saturated NM is a step function 
determined with the Fermi momentum $k^{(\tau)}_F=(3\pi^2\rho_\tau)^{1/3}$ as  
\begin{eqnarray}
 n_{\tau}(k)=\left\{\begin{array}{ccc}
 1 &\mbox{if $k \leqslant k^{(\tau)}_F$} \\
 0 &\mbox{otherwise.} \label{ek3}
\end{array} \right. 
\end{eqnarray}
According to the Landau theory for the infinite Fermi systems \cite{Brown,Migdal}, 
the s/p energy $e_\tau(k)$ in the NM is determined as 
\begin{equation}
e_{\tau}(k)=\frac{\partial E}{\partial n_\tau(k)}=t_{\tau}(k)+U_{\tau}(k)
=\frac{\hbar^2 k^2}{2m_\tau}+U_{\tau}(k), \label{ek4}
\end{equation}
which is the change of the NM energy caused by the removal or addition of a 
nucleon with the momentum $k$. The single-nucleon potential $U_{\tau}(k)$ consists 
of both the HF and rearrangement terms 
\begin{eqnarray}
U_{\tau}(k)&=& U^{\rm (HF)}_{\tau}(k)+U^{\rm (RT)}_{\tau}(k), \label{uk} \\
{\rm where}\ \ U^{\rm (HF)}_{\tau}(k) &=&\sum_{k'\sigma' \tau'}n_{\tau'}(k')
\langle \bm{k}\sigma\tau,\bm{k}'\sigma'\tau'|v_{\rm c}|\bm{k}\sigma\tau,
\bm{k}'\sigma'\tau'\rangle_\mathcal{A} \label{uk1}\\
{\rm and}\ \ \ U^{\rm (RT)}_{\tau}(k) &=&\frac{1}{2}\sum_{k_1\sigma_1\tau_1}
 \sum_{k_2\sigma_2\tau_2}n_{\tau_1}(k_1)n_{\tau_2}(k_2) \nonumber \\
 & & \hskip 2cm\times\left\langle \bm{k}_1\sigma_1\tau_1,\bm{k}_2\sigma_2\tau_2\left|
 \frac{\partial v_{\rm c}}{\partial n_\tau(k)}\right|\bm{k}_1\sigma_1\tau_1,
 \bm{k}_2\sigma_2\tau_2\right\rangle_\mathcal{A}. \label{uk2}
\end{eqnarray}
When the nucleon momentum approaches the Fermi momentum 
$(k\to k^{(\tau)}_F),\ e_\tau(k^{(\tau)}_F)$ determined from 
Eqs.~(\ref{ek4})-(\ref{uk2}) is exactly the Fermi energy given by the 
Hugenholtz - van Hove theorem \cite{HvH}.  Using the transformation \cite{Cze02}
\begin{equation}
\frac{\partial }{\partial n_\tau(k)}\Bigg|_{k\to k^{(\tau)}_F}=
\frac{\partial \rho_\tau}{\partial n_\tau(k^{(\tau)}_F)}
\frac{\partial k^{(\tau)}_F}{\partial \rho_\tau}
\frac{\partial }{\partial k^{(\tau)}_F}=\frac{1}{\Omega}
\frac{\pi^2}{[k^{(\tau)}_F]^2}\frac{\partial }{\partial k^{(\tau)}_F}, 
 \label{ek5}
\end{equation}     
where $\Omega$ is the total volume of the NM in the momentum space, the rearrangement 
term of the s/p potential $U_\tau$ at the Fermi momentum can be obtained \cite{Sat99} as 
\begin{eqnarray}
 U^{\rm (RT)}_\tau(k\to k^{(\tau)}_F) &=& 4\frac{\pi^2}{[k^{(\tau)}_F]^2} 
 \sum_{\tau_1 \tau_2}\frac{\Omega}{2(2\pi)^6} \int\!\!\!\!\int 
 n_{\tau_1}(k_1)n_{\tau_2}(k_2) \hskip 2.5cm \nonumber \\ 
  & \times &\left\langle\bm{k}_1\tau_1,\bm{k}_2\tau_2\left|
 \frac{\partial v_{\rm c}}{\partial k^{(\tau)}_F}\right|
 \bm{k}_1\tau_1,\bm{k}_2\tau_2\right\rangle_\mathcal{A}~d^3k_1 d^3k_2. \label{eRT}
\end{eqnarray}
At variance with the RT part, the HF part of the s/p potential can be readily 
evaluated at any momentum 
\begin{equation}
 U^{\rm (HF)}_\tau(k) = 2\sum_{\tau'}\frac{\Omega}{(2\pi)^3} 
 \int n_{\tau'}(k')\langle\bm{k},\bm{k}'\tau'|v_{\rm c}|
 \bm{k},\bm{k}'\tau'\rangle_\mathcal{A}~d^3k'. \label{eHF} 
\end{equation}     

\begin{figure}[bht]\vspace*{-0.5cm}
\includegraphics[width=0.9\textwidth]{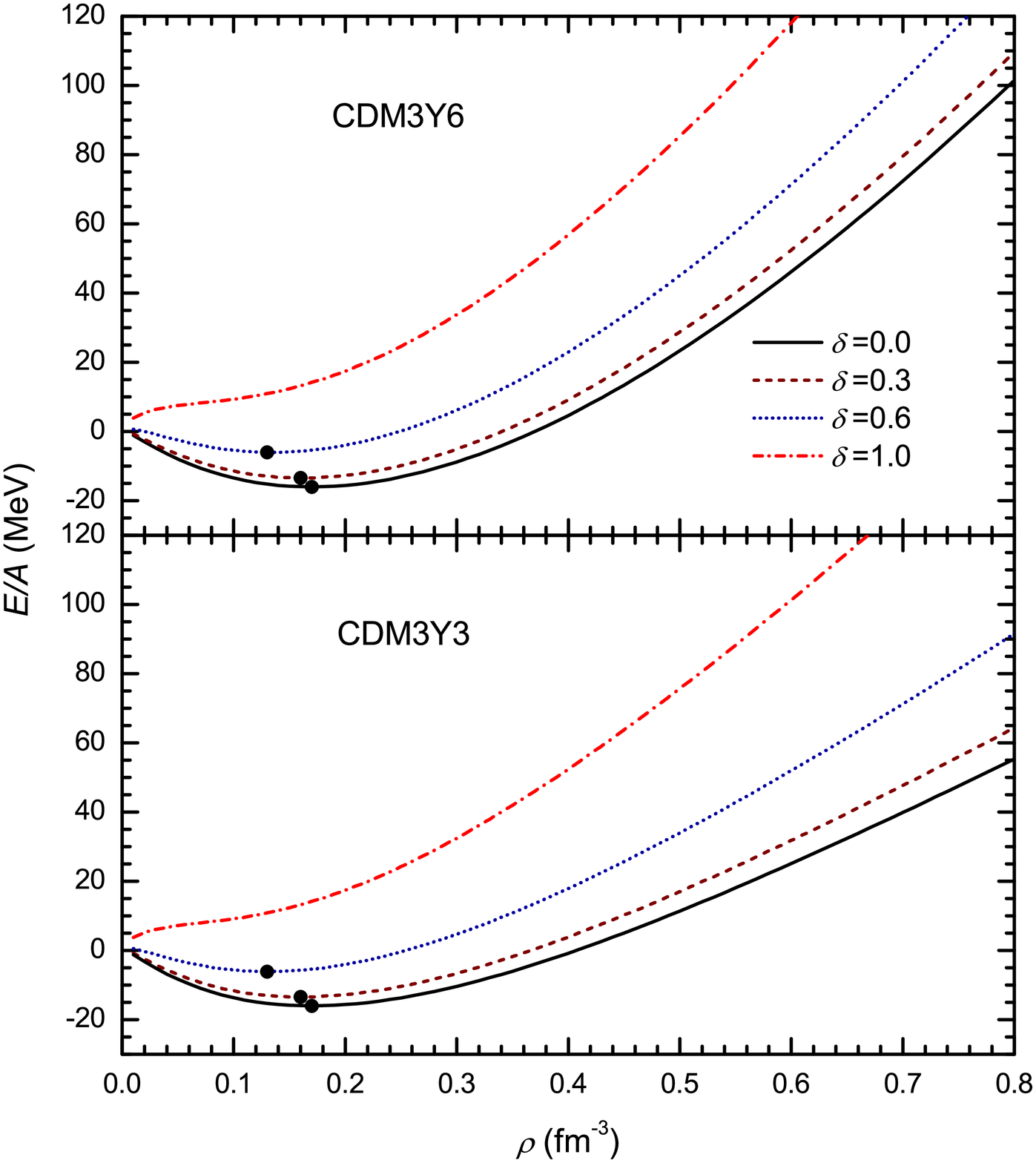}\vspace*{-0.5cm}
 \caption{(Color online) Total NM energy per particle $E/A$ at the different 
n/p asymmetries $\delta$ given by the HF calculation (\ref{ek1})-(\ref{ek2}), 
using the CDM3Y3 (lower panel) and CDM3Y6 (upper panel) interactions with their 
IV density dependence anew determined in the present work. The solid circles are 
the saturation densities of the NM at the different n/p asymmetries.} \label{f1}
\end{figure}
The spin components of plane waves in the HF and RT parts of the s/p potential 
are averaged out, and this results on the spin degeneracy factors 4 and 2 in 
the expressions (\ref{eRT}) and (\ref{eHF}), respectively. We keep in mind that 
the s/p potential is determined consistently at each total NM density 
$\rho$. As a result, $U_\tau$ is a function of the total NM density $\rho$, 
the neutron-proton asymmetry $\delta$, and the nucleon momentum $k$. 
For the spin-saturated NM, only the spin-independent terms of the central NN 
interaction are needed for the determination of the s/p potentials 
(\ref{eRT})-(\ref{eHF}). In the present work, we have used 
two density dependent versions (CDM3Y3 and CDM3Y6) \cite{Kho97} of the M3Y 
interaction based on the G-matrix elements of the Paris NN potential in 
a oscillator basis \cite{An83}. Thus, the central part of the CDM3Yn 
interaction was used in the present HF calculation explicitly as
\begin{equation}
 v^{\rm D(EX)}_{\rm c}(s)=F_0(\rho)v^{\rm D(EX)}_{00}(s) + F_1(\rho)v^{\rm D(EX)}_{01}(s)
 \bm{\tau}_1\cdot\bm{\tau}_2,\ {\rm where}\ s=|\bm{r}_1-\bm{r}_2|. \label{CDM3Y}
\end{equation}  
The radial parts $v^{\rm D(EX)}_{00(01)}(s)$ are kept unchanged as determined 
from the M3Y-Paris interaction \cite{An83}, in terms of three Yukawas
\begin{equation}
 v^{\rm D(EX)}_{00(01)}(s)=\sum_{\nu=1}^3 Y^{\rm D(EX)}_{00(01)}(\nu)
 {{\exp(-R_\nu s)}\over{R_\nu s}}, \label{ykw}
\end{equation}
and the explicit Yukawa strengths and ranges can be found, e.g., in Table~1 
of Ref.~\cite{Kho07}. The density dependence of the interaction (\ref{CDM3Y}) is
assumed to have the same functional form as that introduced first in Ref.~\cite{Kho97}
\begin{equation}
 F_{0(1)}(\rho)=C_{0(1)}[1+\alpha_{0(1)}\exp(-\beta_{0(1)}\rho)+\gamma_{0(1)}\rho].
\label{fden}
\end{equation}
The parameters of the \emph{isoscalar} (IS) density dependence $F_0(\rho)$ were 
determined \cite{Kho97} to reproduce the saturation properties of the symmetric NM and 
give the nuclear incompressibility $K=218$ and 252 MeV with the CDM3Y3 and CDM3Y6 
interactions, respectively. These interactions, especially the CDM3Y6 version, have 
been well tested in numerous folding model analyses of the elastic \nA \cite{Kho02} 
and nucleus-nucleus scattering \cite{Kho07r}, and the charge-exchange scattering to the 
isobar analog states \cite{Kho07,Kho14}. Like in Ref.~\cite{Kho07}, the parameters 
of the \emph{isovector} (IV) density dependence $F_1(\rho)$ were determined  
in the present work to reproduce the BHF results for the IV term of the microscopic 
nucleon OP in the asymmetric NM obtained by Jeukenne, Lejeune and Mahaux 
(JLM) \cite{Je77,Lej80}. Because of the RT included into the extended HF calculation
of the nucleon OP, the parameters obtained for the IV density dependence 
$F_1(\rho)$ are slightly different from those used earlier \cite{Kho07,Kho14}. 
For convenience of the readers who are interested to use the CDM3Yn interaction
in the HF or folding model calculation, the parameters of the density dependence 
are given explicitly in Table~\ref{t1}.  
\begin{table*}
\caption{Parameters of the density dependence (\ref{fden}) of the CDM3Yn interaction. 
The incompressibility $K$ of the symmetric NM, the nuclear symmetry energy $S_0$ and 
its slope $L$ were obtained from the HF results (\ref{es2k}) at the 
saturation density $\rho_0\approx 0.17$ fm$^{-3}$.} \label{t1}\vspace{0.5cm}
\begin{tabular}{|c|c|c|c|c|c|c|c|c|} \hline
Interaction & $i$ & $C_i$ & $\alpha_i$ & $\beta_i$ & $\gamma_i$ & $K$ & $S_0$ & $L$ \\
  &  &  &   & (fm$^3$) & (fm$^3$) & (MeV) & (MeV) & (MeV) \\ \hline
CDM3Y3 & 0 & 0.2985 & 3.4528 & 2.6388 & -1.500 & 218 & - & - \\
       & 1 & 0.2343 & 7.6514 & 9.7494 & 6.6317 & - & 30.1 & 49.6 \\ \hline
CDM3Y6 & 0 & 0.2658 & 3.8033 & 1.4099 & -4.000 & 252 & - & - \\
       & 1 & 0.2313 & 7.6800 & 9.6498 & 6.7202 & - & 30.1 & 49.7 \\ \hline
\end{tabular}
\end{table*} 
			
The HF results for the total energy per particle $E/A$ of the asymmetric NM are 
shown in Fig.~\ref{f1}.
One can see that the saturation density rapidly decreases with the increasing n/p 
asymmetry, and the pure neutron matter ($\delta=1$) is unbound by the (in-medium) 
NN interaction. At the high NM densities, the $E/A$ curves obtained with the CDM3Y6 
interaction are stiffer than those obtained with the CDM3Y3 interaction, and this 
is due to the higher nuclear incompressibility $K$ given by the CDM3Y6 interaction. 
The behavior of the EOS of the asymmetric NM with the increasing n/p asymmetries 
shown in Fig.~\ref{f1} is typical and similar to those observed earlier in the HF 
calculations of the NM using the different types of the in-medium (density dependent) 
NN interaction \cite{Kho96,Tha09,Loa11}. 

Given the parametrization (\ref{CDM3Y}) of the CDM3Y3 and CDM3Y6 interactions, 
the HF part of the s/p potential can be explicitly obtained in terms of the
IS and IV parts as
\begin{eqnarray}
 U_\tau^{\rm (HF)}(\rho,\delta,k) &=& F_0(\rho)U^{\rm (M3Y)}_0(\rho,k)\pm  
  F_1(\rho)U^{\rm (M3Y)}_1(\rho,\delta,k), \label{uHF} \\
	{\rm where}\ 	\ U^{\rm (M3Y)}_0(\rho,k) &=& \rho J^{\rm D}_0 
 + \int A_0(r)v_{00}^{\rm EX}(r)j_0(kr) d^3r, \nonumber \\ 
  {\rm and}\ 	\ U^{\rm (M3Y)}_1(\rho,\delta,k) &=& \rho J^{\rm D}_1\delta
	+\int A_1(r)v_{01}^{\rm EX}(r)j_0(kr) d^3r. \nonumber
\end{eqnarray}
\begin{eqnarray}
{\rm Here}\ A_{0(1)}(r)&=&\rho_n\hat{j_1}(k^{(n)}_F r)\pm\rho_p
\hat{j_1}(k^{(p)}_F r),\ J^{\rm D}_{0(1)}=\int v_{00(01)}^{\rm D}(r)d^3r, \\
{\rm and}\ \hat{j_1}(x)&=& 3j_1(x)/x\ =\ 3(\sin x-x\cos x)/x^3. \nonumber 
\end{eqnarray}
The (-) sign on the right-hand side of Eq.~(\ref{uHF}) pertains to the 
single-proton ($\tau=p$) and (+) sign to the single-neutron ($\tau=n$) potentials.    
Because the original M3Y interaction is momentum independent, the momentum 
dependence of the HF potential (\ref{uHF}) is entirely determined by the exchange 
terms of $U^{\rm (M3Y)}_{0(1)}$. 

Applying the HvH theorem, the RT of the s/p potential is also obtained explicitly 
in terms of the IS and IV parts, but at the Fermi momentum only   
\begin{eqnarray}
 U_\tau^{\rm(RT)}(\rho,\delta, k^{(\tau)}_F) &=& 
  U^{\rm (RT)}_0(\rho,k^{(\tau)}_F)   
 + U^{\rm (RT)}_1(\rho,\delta,k^{(\tau)}_F) \label{uRT} \\
{\rm where}\  U^{\rm (RT)}_0(\rho,k^{(\tau)}_F) &=&\frac{1}{2} 
 \frac{\partial F_0(\rho)}{\partial\rho}\left[\rho^2 J^{\rm D}_0+
\int A_0^2(r)v_{00}^{\rm EX}(r)d^3r \right] \nonumber \\
{\rm and}\  U^{\rm (RT)}_1(\rho,\delta,k^{(\tau)}_F) &=& \frac{1}{2} 
\frac{\partial F_1(\rho)}{\partial \rho}\left[\rho^2 J^{\rm D}_1\delta^2 
+\int A_1^2(r)v_{01}^{\rm EX}(r) d^3r \right]. \nonumber
\end{eqnarray}
One can see from Eq.~(\ref{uRT}) that the RT becomes zero if the density 
independent M3Y interaction is used in the HF calculation of the s/p potential.  
In such a case, the HvH theorem is satisfied already on the HF level \cite{Cze02}.
At a given n/p asymmetry $\delta$, the IV part of the RT is the same for both the 
single-neutron ($\tau=n$) and single-proton ($\tau=p$) potentials, and that 
affects the simple representation of the nucleon OP by Lane \cite{La62,Sat69}, 
where the IV parts of the neutron and proton OP are equal but with the 
\emph{opposite} sign.    

In general, as seen from Eq.~(\ref{uk2}), the RT of the s/p potential should be 
present at the arbitrary nucleon momenta. Microscopically, the momentum dependence 
of the RT was shown, in the BHF calculation of the NM \cite{Mah85,Zuo99,Zuo14}, 
to be due to the higher-order NN correlation, like the second-order diagram 
in the perturbative expansion of the mass operator or the contribution from 
the three-body forces etc. In the finite nuclei, the rearrangement 
effects in the nucleon removal reactions (which have about the same physics origin 
as the RT potential considered here) were shown \cite{Hs75} to be strongly dependent 
on the energy of the stripping reaction, a clear indication of the momentum dependence 
of the RT potential. The question now is whether one can assess the momentum dependence 
of the RT of the s/p potential on the HF level. Making use of the factorized
density dependence of the CDM3Y3 and CDM3Y6 interactions, we suggest in the present 
work a rather simple method to include consistently a momentum dependent RT into the 
s/p potential in the same HF framework. An important constraint of this procedure is 
that adding a realistic momentum dependent RT to the HF potential should improve
the agreement of the calculated nucleon OP in the NM with the empirical data. It has 
been shown \cite{Kho93,Kho95s} that the momentum dependence of the HF potential 
(\ref{uHF}) could account fairly well for the observed energy dependence of the 
nucleon OP after a slight adjustment of the interaction strength at the high energies. 
Therefore, we adopt phenomenologically a momentum dependent RT 
of the s/p potential in the functional form similar to (\ref{uHF}) as  
\begin{equation}
U_\tau^{\rm(RT)}(\rho,\delta, k)= \Delta F_0(\rho)U^{\rm (M3Y)}_0(\rho,k)  
 + \Delta F_1(\rho,\delta)U^{\rm (M3Y)}_1(\rho,\delta,k), \label{uRTk}
\end{equation}
where the (momentum-independent) rearrangement contributions to the IS and IV
density dependences of the CDM3Yn interactions are determined consistently from 
the exact expression (\ref{uRT}) of the RT at the Fermi momentum as   
\begin{equation}
 \Delta F_0(\rho)=\frac{U^{\rm (RT)}_0(\rho,k^{(\tau)}_F)}
 {U^{\rm (M3Y)}_0(\rho,k\to k^{(\tau)}_F)}\ {\rm and}\ 
\Delta F_1(\rho,\delta)=\frac{U^{\rm (RT)}_1(\rho,\delta,k^{(\tau)}_F)}
 {U^{\rm (M3Y)}_1(\rho,\delta,k\to k^{(\tau)}_F)}. \label{dF}
\end{equation}
Consequently, the total s/p potential is determined in the present HF approach as 
\begin{eqnarray}
U_\tau(\rho,\delta,k) &=& U_0(\rho,k)\pm U_1(\rho,\delta,k) \nonumber \\
&=&[F_0(\rho)+\Delta F_0(\rho)]U^{\rm (M3Y)}_0(\rho,k) \nonumber \\
&\pm& [F_1(\rho)\pm \Delta F_1(\rho,\delta)]U^{\rm (M3Y)}_1(\rho,\delta,k), \label{Utotal}
\end{eqnarray}
where (-) sign pertains to $\tau=p$ and (+) sign to $\tau=n$. Thus, the momentum 
dependence of the total s/p potential $U_\tau$ is determined by that of the 
exchange terms of $U^{\rm (M3Y)}_{0(1)}$. Due to the presence of the RT, the 
absolute strength of the IV term of the single-proton potential is not equal 
that of the single-neutron potential. One can see from the expressions 
(\ref{uRTk})-(\ref{Utotal}) that the rearrangement effects actually result 
on a modification of the IS and IV density dependence of the central interaction 
(\ref{CDM3Y}), $v_{\rm c}\to v_{\rm c}+\Delta v_{\rm c}$, so that the total 
s/p potential can be estimated in the standard HF scheme as
\begin{equation}
U_\tau(\rho,\delta,k)=\sum_{k'\sigma' \tau'}n_{\tau'}(k')
\langle \bm{k}\sigma\tau,\bm{k}'\sigma'\tau'|v_{\rm c}+\Delta v_{\rm c}
|\bm{k}\sigma\tau,\bm{k}'\sigma'\tau'\rangle_\mathcal{A}. \label{Uhft}
\end{equation}

\subsection*{Nucleon OP in the NM and the s/p potential at high momenta}
In the NM limit, the nucleon OP is determined as the (mean-field) interaction 
potential between the nucleon incident on the NM at a given energy $E$ and the
bound nucleons in the filled Fermi sea \cite{Kho95s}. In general, the nucleon OP
contains both the IS and IV parts \cite{La62,Sat69} like the total s/p potential 
(\ref{Utotal}). Given a strong dominance of the IS term of the nucleon OP 
\cite{Kho02,Kho09}, one needs first to explore the IS term of the nucleon OP 
predicted by the HF calculation of the NM. Applying a {\em continuous} choice for 
the nucleon s/p potential \cite{Ma91} at the positive energies $E$, we obtain in 
the HF scheme the nucleon OP in the \emph{symmetric} NM \cite{Kho93,Kho95s} as 
\begin{equation}
 U_0(\rho,E)=U^{\rm (HF)}_{\rm IS}(\rho,E)=F_0(\rho)\rho\left[J^D_0+
 \int\hat j_1(k_Fr)j_0\big(k(E,\rho)r\big)v^{\rm EX}_{00}(r)d^3r\right]. \label{Uop}
\end{equation}
Here $k(E,\rho)$ is the (energy dependent) momentum of the incident nucleon 
propagating in the mean field of the nucleons bound in the NM, and is determined as
\begin{equation}
 k(E,\rho)=\sqrt{\frac{2m}{\hbar^2}[E-U_0(\rho,E)]},\ {\rm with}\ E>0. \label{Uopk}
\end{equation}
It is easy to see that $k(E,\rho)>k_F$ and $U^{\rm (HF)}_{\rm IS}$ is just the high 
momentum part of the isoscalar term of the HF potential (\ref{uHF}). Based on the 
above discussion, the total nucleon OP in the NM should also have a contribution 
from the RT added 
\begin{eqnarray}
 U_0(\rho,E)&=& U^{\rm (HF)}_{\rm IS}(\rho,E)+U^{\rm (RT)}_{\rm IS}(\rho,E)
 \nonumber\\ &=& [F_0(\rho)+\Delta F_0(\rho)]
 U^{\rm (M3Y)}_0\big(\rho,k(E,\rho)\big), \label{UopE} 
\end{eqnarray}
where the density dependence $\Delta F_0(\rho)$ of the RT is determined by the 
relation (\ref{dF}). 
\begin{figure}[bht] \vspace*{-0.5cm}
\includegraphics[width=0.9\textwidth]{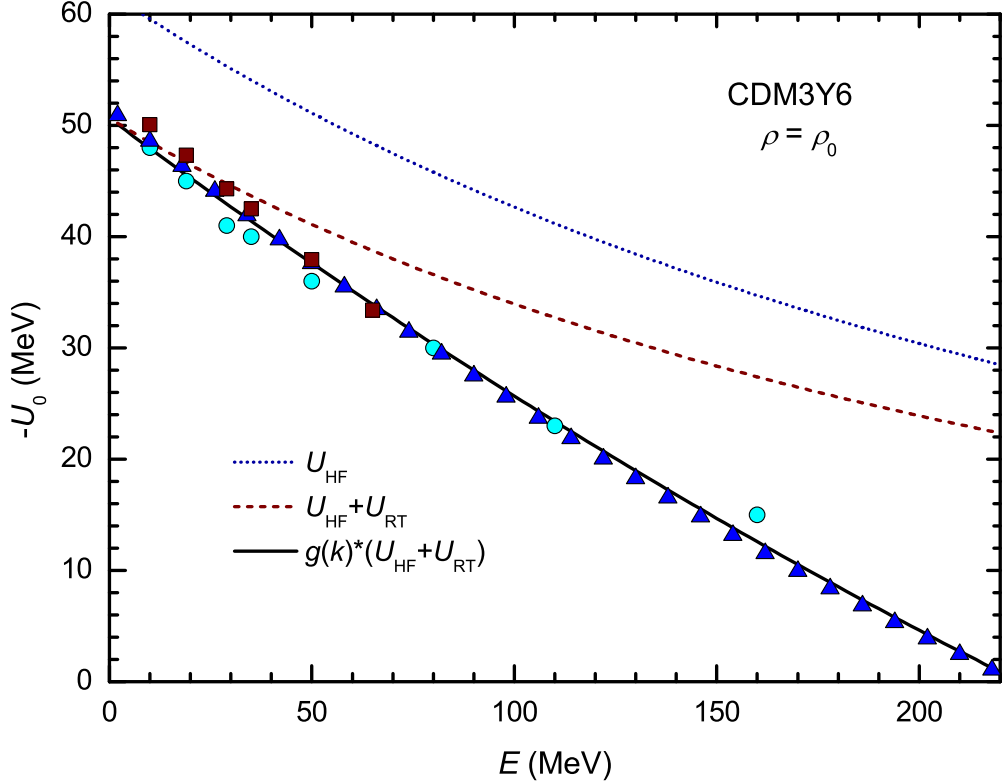}\vspace*{0cm}
 \caption{(Color online) Energy dependence of the nucleon OP in the symmetric NM 
(evaluated at the saturation density $\rho_0$ with and without the RT using the 
CDM3Y6 interaction) in comparison with the empirical data taken from 
Refs.~\cite{BM69} (circles), \cite{Va91} (squares) and \cite{Hama} (triangles). 
The momentum dependent factor $g(k)$ has been iteratively adjusted to the best 
agreement of the calculated nucleon OP (\ref{Uop0}) with the empirical data (solid line).} 
 \label{fUop}
\end{figure}
\begin{figure}[bht] \vspace*{-0.5cm}
\includegraphics[width=0.8\textwidth]{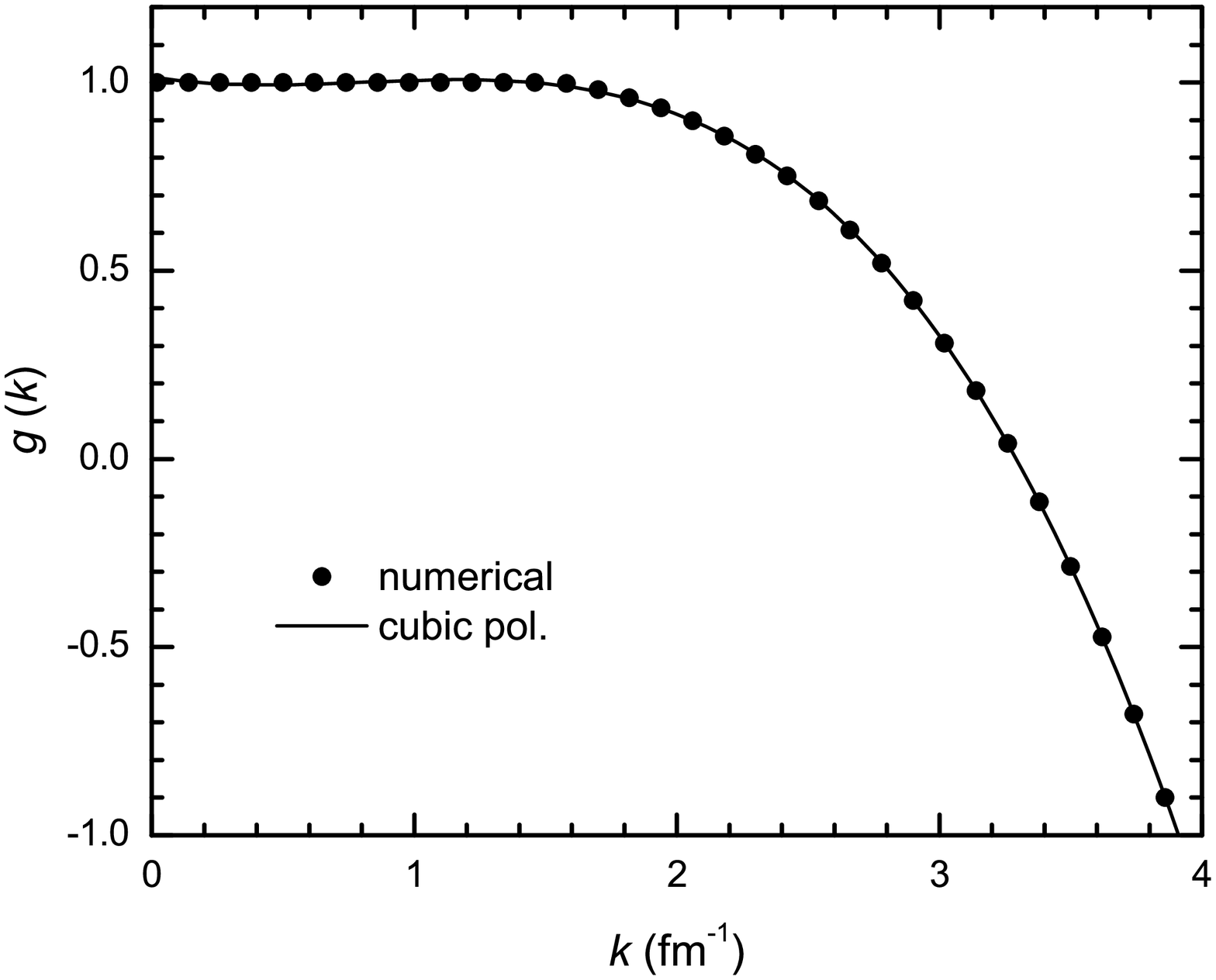}\vspace*{-0.5cm}
 \caption{Momentum dependent scaling factor $g(k)$ obtained with both the CDM3Y3 
and CDM3Y6 interactions from the best (HF+RT) fit of the nucleon OP (\ref{Uop0}) 
to the empirical energy dependence of the nucleon OP \cite{Hama}. The points are 
the numerical results that are well reproduced 
by a cubic polynomial (solid line).} \label{fgk}
\end{figure}

The total nucleon OP (\ref{UopE}) evaluated at the saturation density $\rho_0$ 
of the symmetric NM using the CDM3Y6 interaction are compared with the empirical 
data \cite{BM69,Va91,Hama} in Fig.~\ref{fUop}. Although the inclusion of the RT  
significantly improved the agreement of the calculated $U_0$ with the data at the 
lowest energies, it remains somewhat more attractive at the higher energies in 
comparison with the empirical trend. Such an effect is easily understood in light 
of the microscopic BHF results for the nucleon OP \cite{Ma91}, 
where the energy dependence was shown to come not only from the exchange part, but 
also from the direct part of the microscopic OP because of the energy dependence 
of the Brueckner G-matrix. That is the reason why a slight linear energy dependence 
has been introduced into the CDM3Y6 interaction \cite{Kho97,Kho02}, in terms of the 
$g(E)$ factor. To be consistent with the momentum dependence of the s/p potential 
under study, instead of the $g(E)$ factor, we scale in the present work the CDM3Yn 
interaction (\ref{CDM3Y}) by a momentum dependent function $g\big(k(E,\rho)\big)$, 
and iteratively adjust its strength to the best agreement of the (HF+RT) nucleon OP 
obtained at the saturation density $\rho_0$ with the empirical data, as shown in 
Fig.~\ref{fUop}. As a result,
\begin{equation}
U_0(\rho,E)=g\big(k(E,\rho)\big)[F_0(\rho)+\Delta F_0(\rho)]
U^{\rm (M3Y)}_0\big(\rho,k(E,\rho)\big), \label{Uop0} 
\end{equation}
where $k(E,\rho)$ is determined self-consistently from $U_0(E,\rho)$ by 
Eq.~(\ref{Uopk}). At variance with the $g(E)$ factor fixed by the incident energy 
\cite{Kho97,Kho02}, $g\big(k(E,\rho)\big)$ is now a momentum dependent function 
(see Fig.~\ref{fgk}), carrying the important signature of the momentum 
dependence of the nucleon mean-field potential. Numerically, the obtained $g(k)$ 
function is nearly identical for both the CDM3Y3 and CDM3Y6 interactions, and it can be 
considered as the explicit momentum dependence of the CDM3Yn interaction that allows 
the incident nucleon to feel the nucleon mean-field potential during its interaction 
with the nucleons bound in the NM. In this sense, such a momentum dependence is 
of a similar nature as the momentum dependence of the G-matrix in the microscopic 
BHF study of NM, which is determined self-consistently through the momentum dependence 
of the s/p energies embedded in the denominator of the Bethe-Goldstone equation 
\cite{Bom91,Zuo99}. The technical difference here is that the $k$-dependence of $g(k)$ 
has been determined from the best fit of the calculated s/p potential (\ref{Uop0}) 
at positive energies with the observed energy dependence of the nucleon OP. It can be 
seen in Fig.~\ref{fgk} that $g(k)$ becomes smaller unity at $k\gtrsim 1.6$ fm$^{-1}$ 
only. Consequently, the obtained $g(k)$ function is used further in the extended HF 
calculation to adjust the high-momentum part of the (HF+RT) s/p potential. 
  
\begin{figure}[bht] \vspace*{-0.5cm}
\includegraphics[width=0.8\textwidth]{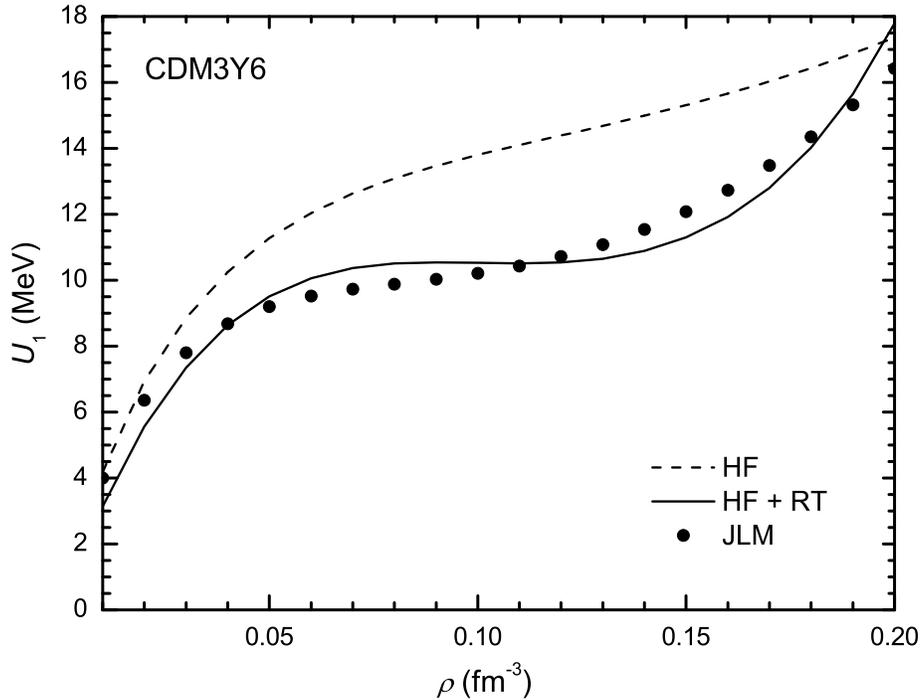}\vspace*{0cm}
 \caption{Density dependence of the IV part of the neutron OP in the pure neutron 
matter at the energy $E=1$ MeV, evaluated  with and without the RT. The parameters 
(\ref{fden}) of the IV density dependence $F_1(\rho)$ of the CDM3Y6 interaction 
have been iteratively adjusted to give the best agreement of the HF+RT 
result with that of the BHF calculation by the JLM group (solid circles) 
\cite{Je77,Lej80}.} \label{fUop1}
\end{figure}

Situation is quite different concerning the IV term of the nucleon OP because
there are no systematic (energy dependent) empirical data available, like those 
discussed above for the IS potential. However, it is well established from 
numerous optical model analyses of the elastic nucleon scattering that the 
absolute strength of the IV term of the nucleon OP is much weaker than 
that of the IS \cite{Kho02,Kho07,Va91}, and the energy dependence of the nucleon
OP is dominantly determined by that of the IS term \cite{Va91}. Consequently, the 
momentum dependent function $g(k)$ determined above for the IS part should be a 
reasonable approximation for the IV part of the nucleon OP in the HF+RT calculation
\begin{equation}
 U_1(\rho,\delta,E)=g\big(k(E,\delta,\rho)\big)[F_1(\rho)+\Delta F_1(\rho,\delta)]
 U^{\rm (M3Y)}_1\big(\rho,\delta,k(E,\delta,\rho)\big), \label{Uop1} 
\end{equation}
where $k(E,\delta,\rho)$ is determined self-consistently from the total nucleon OP 
in the asymmetric NM as 
\begin{equation}
 k(E,\delta,\rho)=\sqrt{\frac{2m}{\hbar^2}[E-U_0(\rho,E)\mp U_1(\rho,\delta,E)]}.  
 \label{Uopk1}
\end{equation}
Thus, the NN interaction in the $\tau.\tau$ channel of the central force (\ref{CDM3Y}) 
is also influenced by the nucleon mean-field potential through the momentum dependent 
$g(k)$ function. Like in Refs.~\cite{Kho07,Kho14}, parameters of the IV density 
dependence $F_1(\rho)$ of the M3Y-Paris interaction have been iteratively adjusted 
in the HF+RT calculation to achieve a good agreement of the IV potential (\ref{Uop1}) 
with the microscopic nucleon OP in the asymmetric NM given by the BHF calculation 
done by the JLM group \cite{Je77,Lej80}, with the contribution of the RT properly 
included \cite{Je77a}. 
At variance with the previous studies where $F_1(\rho)$ has been determined 
separately at each considered energy \cite{Kho07,Kho14}, we have used in  
the present work a unique set of the parameters of $F_1(\rho)$ which were 
determined from the best fit of the IV potential (\ref{Uop1}) to the corresponding 
JLM results at the lowest energies \cite{Lej80}, where $g(k)\approx 1$ 
(see Fig.~\ref{fUop1}). Together with the IS density dependence $F_0(\rho)$ 
determined earlier \cite{Kho97}, the newly determined IV density dependence 
$F_1(\rho)$ of the CDM3Yn interactions have been used in the present work to calculate 
the total NM energy and the single-nucleon potential. The absolute strength 
of $F_1(\rho)$ in the extended HF calculation was scaled by a factor of 1.3, which 
was found necessary in the folding model analysis of the $(p,n)$ scattering to the 
isobar analog states \cite{Kho07}. As a result, the nuclear symmetry energy 
$S(\rho_0)$ at the saturation density given by the present HF calculation 
is very close to the empirical value of about 30 MeV (see also next Section). The 
final values of the parameters of $F_0(\rho)$ and $F_1(\rho)$ are given with the most 
important NM properties in Table~\ref{t1}. 

\begin{figure}[bht]\vspace*{-0.5cm}
\includegraphics[width=0.8\textwidth]{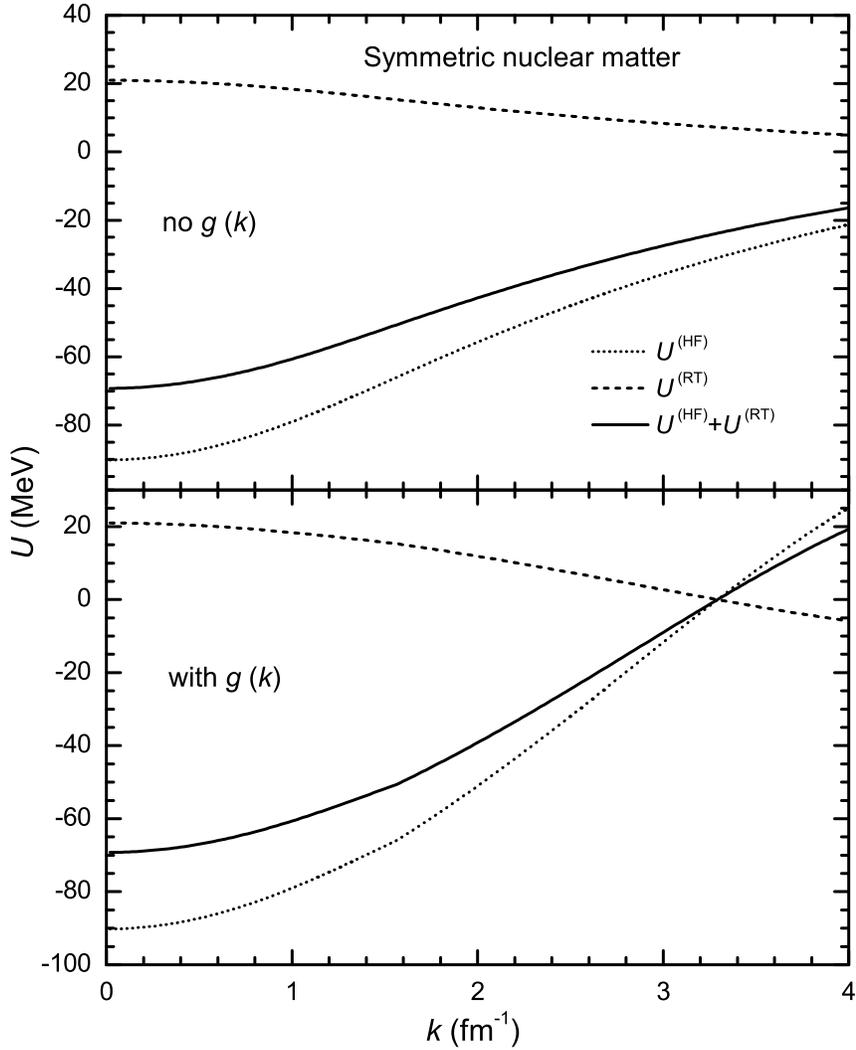}\vspace*{-0.5cm}
 \caption{Momentum dependence of the total s/p potential in the symmetric NM
at the saturation density $\rho_0$, with the explicit contributions  
from the RT and HF parts. The upper panel shows the results of the extended 
HF calculation (\ref{UopE}), and the lower panel shows the total s/p potential  
(\ref{Uop0}), with the high-momentum part corrected by the $g(k)$ function 
determined from the observed energy dependence of the nucleon OP.} \label{fUk}
\end{figure}
\begin{figure}[bht]\vspace*{-0.5cm}
\includegraphics[width=0.8\textwidth]{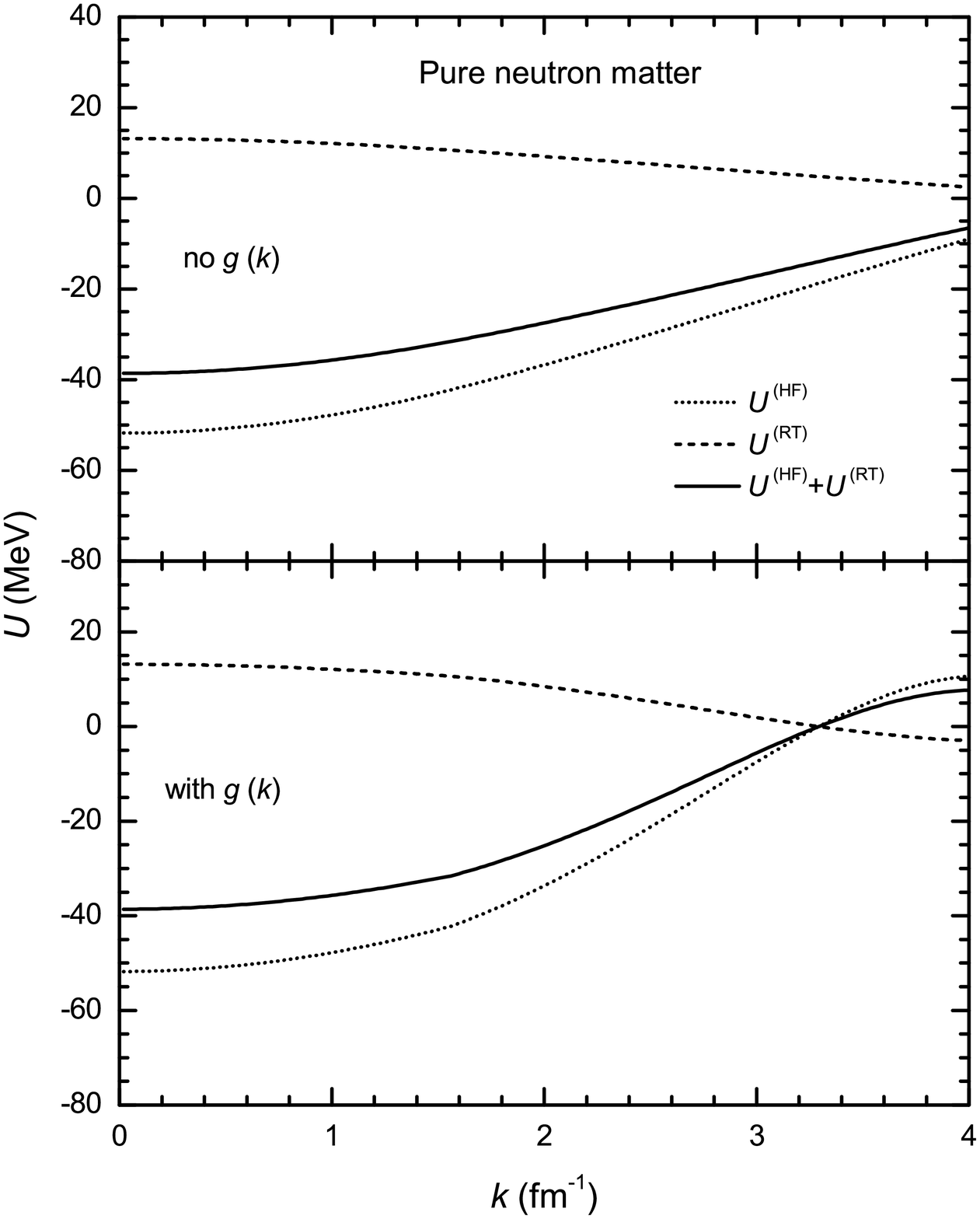}\vspace*{-0.5cm}
 \caption{The same as Fig.~\ref{fUk} but for the total s/p potential in the 
pure neutron matter.} \label{fUk1}
\end{figure}

The total (density- and momentum dependent) s/p potential is now determined as 
\begin{equation}
 U_\tau(\rho,\delta,k)=g(k)[U_0(\rho,k)\pm U_1(\rho,\delta,k)], \label{Utotalg} 
\end{equation}
where $U_0(\rho,k)$ and $U_1(\rho,\delta,k)$ are determined by using Eq.~(\ref{Utotal})
and the same function $g(k)$ as that shown in Fig.~\ref{fgk}. The total s/p potentials 
obtained at the saturation density $\rho_0\approx 0.17$ fm$^{-3}$ from the extended HF 
calculation (\ref{Uhft}) of the symmetric NM and the pure neutron matter using 
the CDM3Y6 interaction are shown in Figs.~\ref{fUk} and \ref{fUk1}, respectively. 
One can see that the RT is largest at the nucleon momenta close to zero, which 
correspond to nucleons deeply bound in the NM ($k\ll k_F$). The RT steadily decreases 
with the increasing nucleon momentum, and the decrease of the RT becomes faster 
when the high-momentum part of the s/p potential is scaled by the $g(k)$ function
determined [see Eqs.~(\ref{Uop0}) and (\ref{Uop1})] to reproduce the observed 
energy dependence of the nucleon OP. In this case, the s/p potential reaches zero 
and changes sign at the momentum $k\approx 3.3$ fm$^{-1}$ that corresponds to the 
nucleon OP at the incident energy around 220 MeV, where the empirical 
(Schr\"odinger-equivalent) $U_{\rm op}$ was shown to become repulsive \cite{Hama}.   
Such a momentum dependence of the s/p potential agrees well with that predicted 
by the microscopic BHF calculation (see, e.g., Figs. 6 and 7 of Ref.~\cite{Zuo99}). 
A decrease of the RT with the increasing nucleon momentum (or energy) also 
agrees with the observed energy dependence of the rearrangement effect in the 
nucleon removal reactions \cite{Hs75}. In the relative strength, the RT contributes 
to about $20\sim 30$ \% of the total strength of the s/p potential at $\rho=\rho_0$ 
over a wide range of the nucleon momentum and is, therefore, a very clear 
manifestation of the HvH theorem \cite{HvH}.      

\begin{figure}[bht]\vspace*{-1cm}
\includegraphics[width=0.8\textwidth]{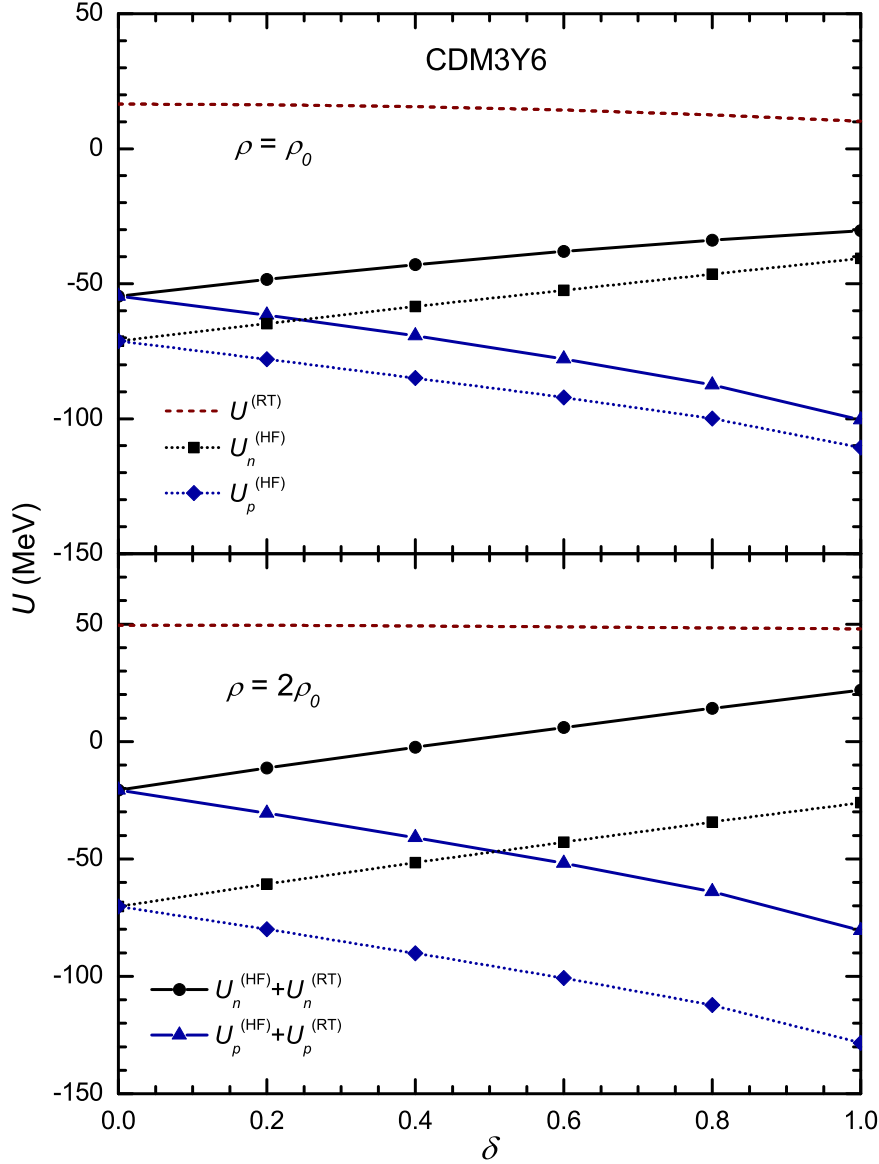}\vspace*{-0.5cm}
\caption{(Color online) Contributions of the RT and HF parts to the total
neutron and proton s/p potentials evaluated at the different n/p asymmetries 
$\delta$ and nucleon momenta $k=k^{(\tau)}_F$, using the CDM3Y6 interaction. 
The results obtained at the saturation density $\rho=\rho_0$ are shown in the 
upper panel, and those obtained at the density $\rho=2\rho_0$ are shown 
in the lower panel.} \label{fUdelta}
\end{figure}
The dependence of the neutron and proton s/p potentials on the n/p asymmetry 
(with the explicit contributions from the RT and HF parts) is shown in 
Fig.~\ref{fUdelta}, and one can see that the RT is the same for both the neutron 
and proton s/p potentials. In the present extended HF scheme, such an equality is 
exactly obtained from the RT given by the HvH theorem (\ref{uRT}). At a given NM 
density, the RT decreases slightly with the increasing n/p asymmetry $\delta$. 
As can bee seen from the lower panel of Fig.~\ref{fUdelta}, the repulsive 
contribution of the RT becomes much stronger at the high density $\rho=2\rho_0$, 
with the relative strength up to 70 \% of the HF term. Such a behavior of the
RT is well expected, given the higher-order NN correlations and the three-body
forces as the physics origin of the RT \cite{Mah85,Zuo99}, which become much
more substantial with the increasing NM density. The results of our extended HF 
calculation shown in Fig.~\ref{fUdelta} also agree well with those of the recent 
BHF calculation of the asymmetric NM by Vida\~na \cite{Vi13}, using the Argon NN 
interaction ($V_{18}$ version \cite{Wir95} with the Urbana three-body force). 

To conclude this section, a simple and consistent method has been developed to 
account effectively for the momentum dependence of the RT of the s/p potential 
in an extended HF calculation using the CDM3Yn density dependent interactions 
(\ref{CDM3Y}), based on the exact expression of the RT given by the HvH theorem 
at each NM density and the empirical energy dependence of the nucleon OP 
observed over a wide range of energies.

\section{Single-nucleon potential and the symmetry energy}
\label{sec2}
Given the importance of the nuclear symmetry energy for the nuclear astrophysics
studies, especially, its vital role in the determination of the EOS of the asymmetric 
NM \cite{Hor14,Ba08}, we focus in the present section on the connection of the 
s/p potential in the NM with the nuclear symmetry energy, which has been widely
discussed in Refs.~\cite{Xu14,Zuo14}. 
The nuclear symmetry energy $S(\rho)$ is normally defined with an expansion of the 
total NM energy per particle over the n/p asymmetry $\delta$ as  
\begin{equation}
\frac{E}{A}(\rho,\delta)=\frac{E}{A}(\rho,\delta=0)+ 
S(\rho)\delta^2+O(\delta^4)+... \label{es1}
\end{equation}
For $\delta<1$, the contribution of $O(\delta^4)$ and the higher-order terms 
in Eq.~(\ref{es1}) was proven to be quite small \cite{Kho96,Zuo99} and 
can be neglected (the so-called \emph{parabolic} approximation). For the pure
neutron matter ($\delta=1$) the convergence of the series (\ref{es1}) is slower, 
and the parabolic law becomes less accurate. Therefore, a more general definition 
of the nuclear symmetry energy as the energy required per particle to change the 
symmetric NM into the pure neutron matter is also used in the mean-field studies
\begin{equation}
S(\rho)=\frac{E}{A}(\rho,\delta=1)-\frac{E}{A}(\rho,\delta=0). \label{es2}
\end{equation}
\begin{figure}[bht] \vspace*{-0.5cm}
\includegraphics[width=0.9\textwidth]{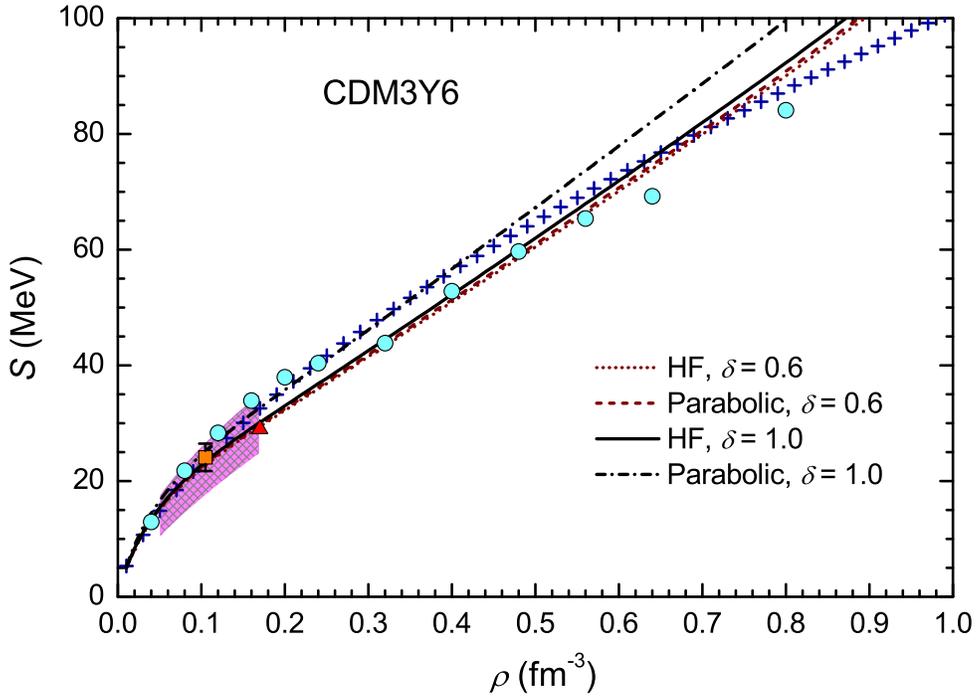}\vspace*{-1.5cm}
 \caption{(Color online) Nuclear symmetry energy $S(\rho)$ at the n/p 
asymmetries $\delta=0.6$ and 1 obtained with the CDM3Y6 interaction, using 
the parabolic approximation (\ref{es1}) and general definition (\ref{es2}). 
The shaded (magenta) region marks the empirical boundaries implied by the analysis 
of the isospin diffusion data and the double ratio of the neutron and proton spectra 
observed in the HI collisions \cite{Ono03,Tsa09}. The square is the empirical $S$ 
value implied by the structure study of the GDR \cite{Tri08}, and the triangle is 
that established from the analysis of the terrestrial nuclear physics experiments 
and astrophysical observations \cite{Li13}. The circles and crosses are results 
of the ab-initio calculations by Akmal {\it et al.} \cite{Ak98} and Gandolfi {\it et al.} 
\cite{Gan10}, respectively.} \label{fSpa}
\end{figure}
The nuclear symmetry energy $S(\rho)$ can also be expanded \cite{Tsa09} around 
$\rho_0$ as
\begin{equation}
 S(\rho)=S_0+\frac{L}{3}\left(\frac{\rho-\rho_0}{\rho_0}\right) +
 \frac{K_{\rm sym}}{18}\left(\frac{\rho-\rho_0}{\rho_0}\right)^2+ ... \label{es2k}
\end{equation}
where $L$ and $K_{\rm sym}$ are the slope and curvature parameters of the
symmetry energy at $\rho_0$. While the curvature parameter $K_{\rm sym}$ is still
poorly known, it has been shown recently by Li and Han \cite{Li13} that 
quite a robust constraint for both $S_0$ and $L$ values has been established 
based on several tens analyses of the terrestrial nuclear physics experiments 
and astrophysical observations, which give $S_0\approx 31.6\pm 2.7$ MeV and 
$L\approx 58.9\pm 16.0$ MeV. With the parameters of the IV density dependence 
of the CDM3Yn interaction anew determined in the present work, the values 
of $S_0\approx 30.1$ MeV and $L\approx 49.7$ MeV given by the HF calculation
are well within this empirical range.
    
Applying the HvH theorem to calculate the neutron and proton energies (\ref{ek4})
at the corresponding Fermi momenta, one obtains \cite{Xu14,Zuo14} the nuclear 
symmetry energy directly from the difference between the neutron and proton 
Fermi energies as  
\begin{equation}
 t_n(k^{(n)}_F)-t_p(k^{(p)}_F)+U_n(\rho,\delta,k^{(n)}_F)-
 U_p(\rho,\delta,k^{(p)}_F)=4S(\rho)\delta+O(\delta^3)+... \label{es3}
\end{equation}
Thus, Eq.~(\ref{es3}) is simply the parabolic law in the s/p energy representation. 
As discussed in Sect.~\ref{sec1}, at given $\rho$ and $\delta$ values the 
contribution of the RT to the s/p potential (\ref{uRT}) is the same for both the 
neutron and proton s/p potentials. Therefore, the RT contribution to the nuclear 
symmetry energy $S(\rho)$ through the difference between the neutron and proton s/p 
potentials in Eq.~(\ref{es3}) is canceled out. The same conclusion was also drawn in 
the recent BHF study of asymmetric NM by Vida\~na \cite{Vi13}. Using the 
single-nucleon potentials given by the present HF calculation, we obtained exactly
the same $S(\rho)$ from both Eq.~(\ref{es3}) and the expansion (\ref{es1}), neglecting
the higher-order terms $O(\delta^3)$ and $O(\delta^4)$, respectively. These results
are compared with those given by the general definition (\ref{es2}) in Fig.~\ref{fSpa}.
One can see that the parabolic approximation is reasonable only for $\delta<1$, 
and it becomes poorer for pure neutron matter. As a result, the higher-order terms 
on the right-hand side of Eqs.~(\ref{es1}) and (\ref{es3}) need to be taken into 
account for very neutron-rich matter, as done recently by Chen {\it et al.} \cite{Che12}.    

\begin{figure}[bht] \vspace*{-0.5cm}
\includegraphics[width=0.8\textwidth]{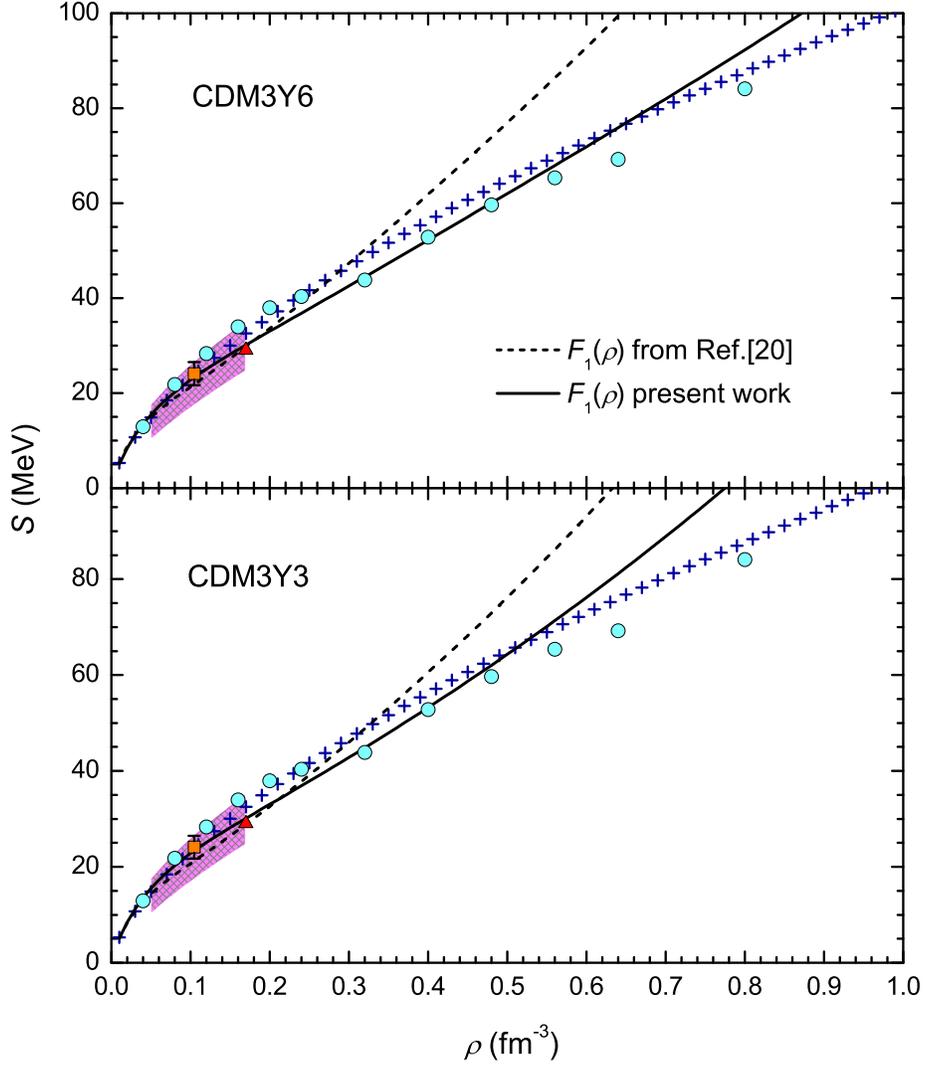}\vspace*{-0.5cm}
 \caption{(Color online) Nuclear symmetry energy $S(\rho)$ given by  
the HF calculation using the new parameters of the IV density dependence 
$F_1(\rho)$ of the CDM3Y6 (upper panel) and CDM3Y3 (lower panel) interactions. 
These same HF results obtained with the old parameters of $F_1(\rho)$ determined in 
Ref.~\cite{Kho07} are shown as the dashed lines. Other symbols are 
the same as in Fig.~\ref{fSpa}} \label{fEsym}
\end{figure}
The main method to probe $S(\rho)$ obtained with a chosen in-medium NN interaction 
is to probe this interaction in the analysis of heavy-ion (HI) collisions \cite{Ono03,Tsa09} 
or in the structure studies of nuclei with large neutron excess \cite{Tri08,Fur02}. 
Based on the constraints implied by such studies, extrapolation is often made to study 
the low- and high-density behavior of nuclear symmetry energy. For the illustration, we 
have compared in Fig.~\ref{fSpa} the HF results given by the CDM3Y6 interaction for 
$S(\rho)$ with the empirical data  \cite{Ono03,Tsa09,Tri08,Li13,Fur02} 
and the results of the ab-initio calculations of the asymmetric NM by Akmal {\it et al.} 
\cite{Ak98} and Gandolfi {\it et al.} \cite{Gan10}. Around the saturation density 
$\rho_0$ the symmetry energy $S_0$ given by the HF calculation is in a very good 
agreement with the empirical value \cite{Fur02,Li13}. In the low-density 
region ($\rho\approx 0.3 \sim 0.6 \rho_0$) the empirical boundaries for $S(\rho)$ 
deduced from the analysis of the isospin diffusion data and double ratio of neutron 
and proton spectra data of HI collisions \cite{Ono03,Tsa09} do enclose our HF result.
At the density $\rho\approx 0.1$ fm$^{-3}$, the HF result also agrees well with the 
empirical value deduced from the structure study of the Giant Dipole Resonance (GDR) 
in heavy nuclei \cite{Tri08}. At supranuclear densities, where the reliable empirical
data are still absent, our HF results follow closely those of the ab-initio 
calculations \cite{Ak98,Gan10}.  

\begin{figure}[bht] \vspace*{-0.5cm}
\includegraphics[width=0.8\textwidth]{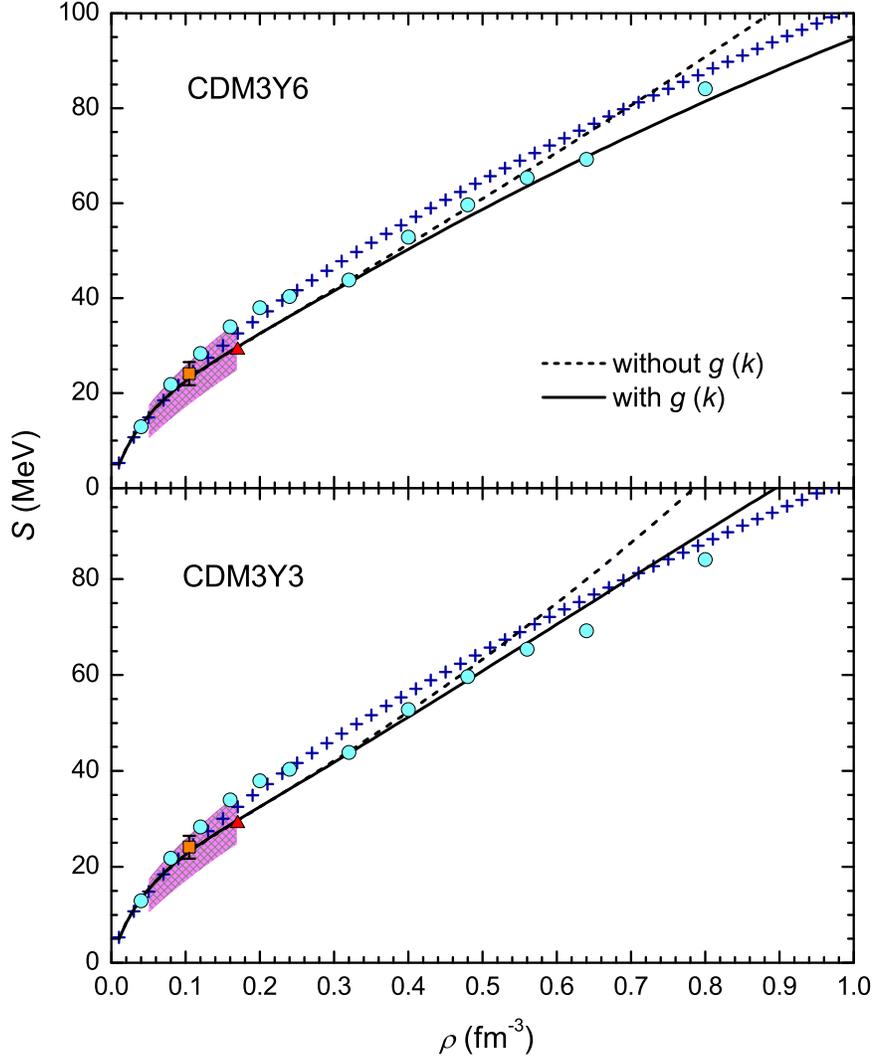}\vspace*{-0.5cm}
 \caption{(Color online) Nuclear symmetry energy $S(\rho)$ obtained at the n/p 
asymmetry $\delta<1$ from the (HF+RT) s/p energies with the CDM3Y6 (upper panel) 
and CDM3Y3 (lower panel) interactions, using the parabolic approximation 
(\ref{es3}). The solid (dashed) curves show the results obtained with (without) 
the modification of the high-momentum part of the s/p potential by the $g(k)$ 
function determined from the observed energy dependence of the nucleon OP. 
Other symbols are the same as in Fig.~\ref{fSpa}} \label{fEsymk}
\end{figure}
Although the explicit contribution of the RT to the nuclear symmetry energy, through 
the difference between the neutron and proton s/p potentials (\ref{es3}), is canceled 
out. The vital role of the rearrangement effects in the HF calculation of the nuclear 
symmetry energy is well illustrated in Fig.~\ref{fEsym}, where $S(\rho)$ values 
obtained from the HF calculation using the new parameters of the IV density dependence 
$F_1(\rho)$ of the CDM3Yn interactions are compared with those obtained with the 
old parameters of $F_1(\rho)$, determined in Ref.~\cite{Kho07} by fitting the IV part 
of the (HF only) s/p potential to the IV part of the microscopic JLM potential. 
One can see in Fig.~\ref{fEsym} that $S(\rho)$ given by the old IV density dependence 
agrees with the empirical data and ab-initio results at NM densities up to about 
2$\rho_0$ only. Using the new parameters of $F_1(\rho)$ obtained in the present work 
by fitting the IV part of the (HF+RT) s/p potential to the JLM potential 
(see Fig.~\ref{fUop1}), the calculated $S(\rho)$ values now follow closely
those given by the ab-initio calculations over a wider range of the NM densities, 
up to $\rho\approx 4\rho_0$.    

With the Fermi momentum $k_F$ becoming larger at high NM densities, a realistic 
momentum dependence of the s/p potentials should be helpful in constraining 
the nuclear symmetry energy $S(\rho)$ at the supranuclear densities, based
on the relation (\ref{es3}). The symmetry energies $S(\rho)$ given by the 
difference between the neutron- and proton s/p energies (\ref{es3}) at the n/p 
asymmetries $\delta<1$ are shown in Fig.~\ref{fEsymk}, with and without the 
modification of the high-momentum part of the (HF+RT) nucleon OP by the $g(k_F)$ 
function determined from the observed energy dependence of the nucleon OP. One can see 
that the modification of the high-momentum tail of the (HF+RT) s/p potential results 
on a slightly softer slope of the nuclear symmetry energy at the high NM densities that 
leads, in turn, to a good agreement of the HF results with those of the ab-initio 
calculations over a much wider range of the NM density, up to $\rho\approx 5\sim 6 \rho_0$. 
This result is, thus, complementary to the recent efforts by Li {\it et al.} 
\cite{XLi13} to determine the nuclear symmetry from the optical model analysis 
of the elastic neutron-nucleus scattering over a wide range of energies.      

\section{Neutron-proton effective mass splitting}
\label{sec3}
As discussed in Sec.~\ref{sec1}, due to the finite range of the Yukawa functions 
in the radial part of the CDM3Yn interaction (\ref{CDM3Y}), the s/p potential 
depends explicitly on the nucleon momentum $k$ through its exchange term that 
implies a \emph{nonlocal} single-nucleon potential in the coordinate 
space. At the high nucleon momenta, the momentum dependence of the s/p potential
is further modified by the $g(k)$ function implied by the observed energy dependence
of the nucleon OP. An important quantity associated with the momentum dependence
of the nucleon s/p potential is the nucleon effective mass $m^*_\tau$, defined 
within the nonrelativistic mean-field formalism as
\begin{equation}
\frac{m^*_\tau(\rho,\delta)}{m}=\Biggl[1+\frac{m}{\hbar^2k^{(\tau)}_F}
\frac{\partial U_\tau(\rho,\delta,k)}{\partial k}\Bigg|_{k^{(\tau)}_F}\Biggr]^{-1}, 
 \label{es4}
\end{equation}
where $m$ is the free nucleon mass. Thus, the nucleon effective mass $m^*_\tau$ 
describes the nonlocality of the mean-field potential felt by a nucleon propagating 
through the nuclear medium. 
\begin{figure}[bht] \vspace*{-0.5cm}
\includegraphics[width=0.8\textwidth]{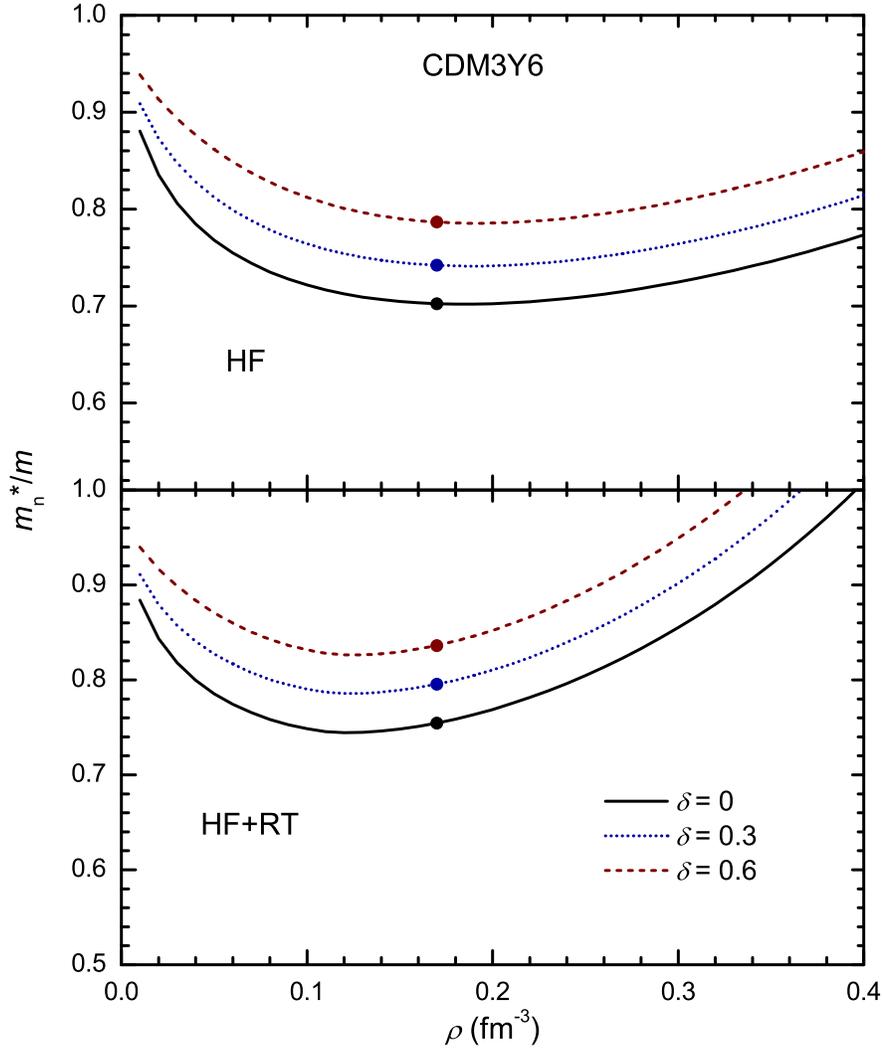}\vspace*{-0.5cm}
 \caption{(Color online) Density dependence of the neutron effective mass 
(\ref{es4}) at the n/p asymmetry $\delta=0,\ 0.3$ and 0.6, obtained from the 
s/p potential given by the CDM3Y6 interaction. The HF and HF+RT results 
are shown in the upper and lower panels, respectively. The solid circles
are the effective-mass values determined at the saturation density $\rho_0$.} 
\label{fEffm}
\end{figure}
\begin{figure}[bht] \vspace*{-0.5cm}
\includegraphics[width=0.8\textwidth]{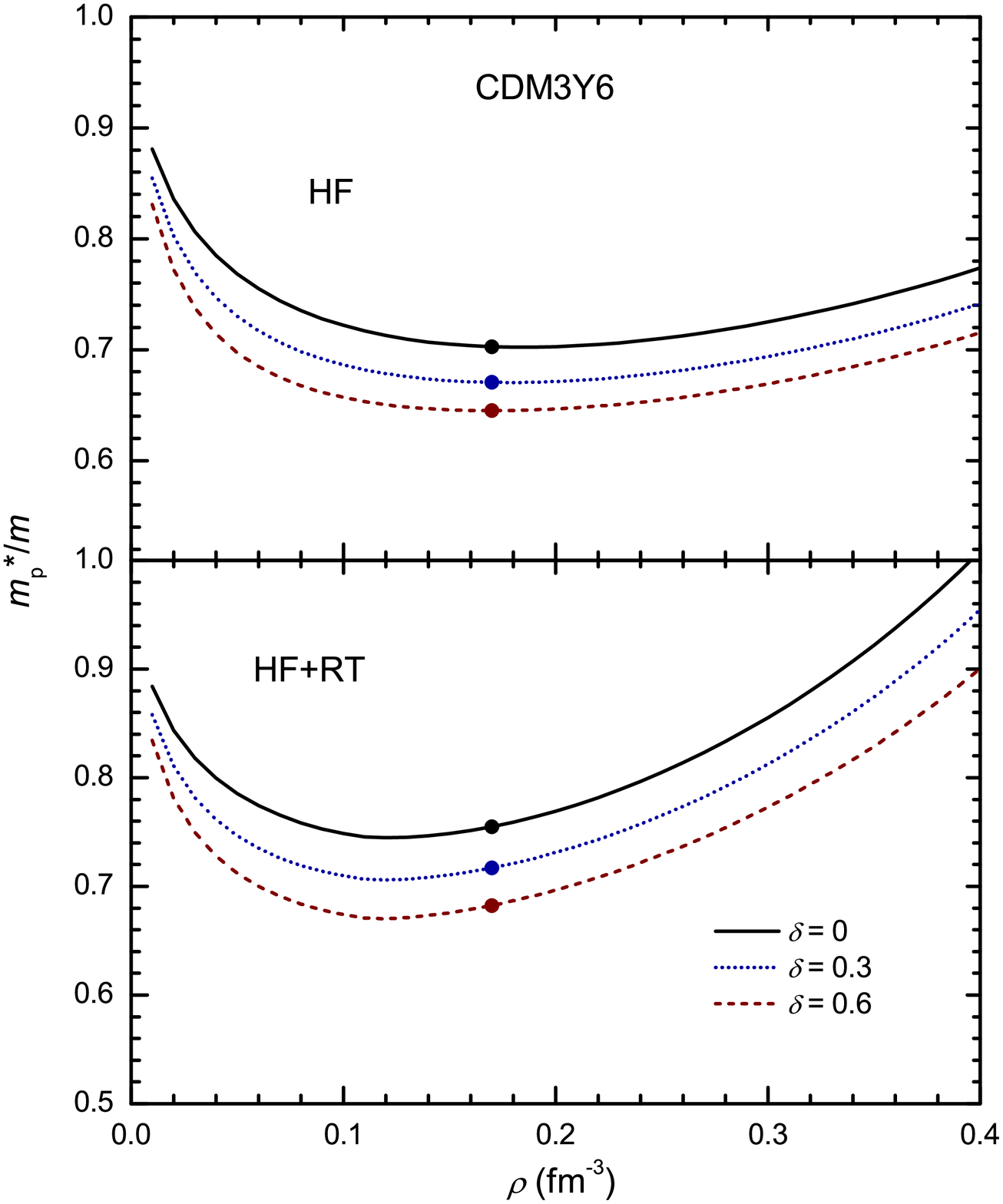}\vspace*{-0.5cm}
 \caption{(Color online) The same as Fig.~\ref{fEffm} but for the proton
effective mass.}  
\label{fEffm1}
\end{figure}
As such, the nucleon effective mass is directly linked to many nuclear physics 
phenomena, like the dynamics of HI collisions, damping of giant resonances, 
temperature profile of the hot stellar objects and neutrino emission therefrom 
\cite{Bal14}. For the asymmetric NM, the relative difference between the neutron and 
proton effective masses
\begin{equation}
m^*_{n-p}(\rho,\delta)=\frac{m^*_n(\rho,\delta)-m^*_p(\rho,\delta)}{m} \label{es5}
\end{equation}
is widely discussed \cite{Li13,Li15} as the \emph{neutron-proton effective 
mass splitting}, which is closely related to the nuclear symmetry energy $S(\rho)$ 
and its slope parameter $L$ \cite{Li13}. The n/p effective mass splitting was also 
suggested to affect the neutron/proton ratio during the stellar evolution, and the 
cooling of protoneutron stars etc. \cite{Li15}.

\begin{figure}[bht] \vspace*{-0.5cm}
\includegraphics[width=0.8\textwidth]{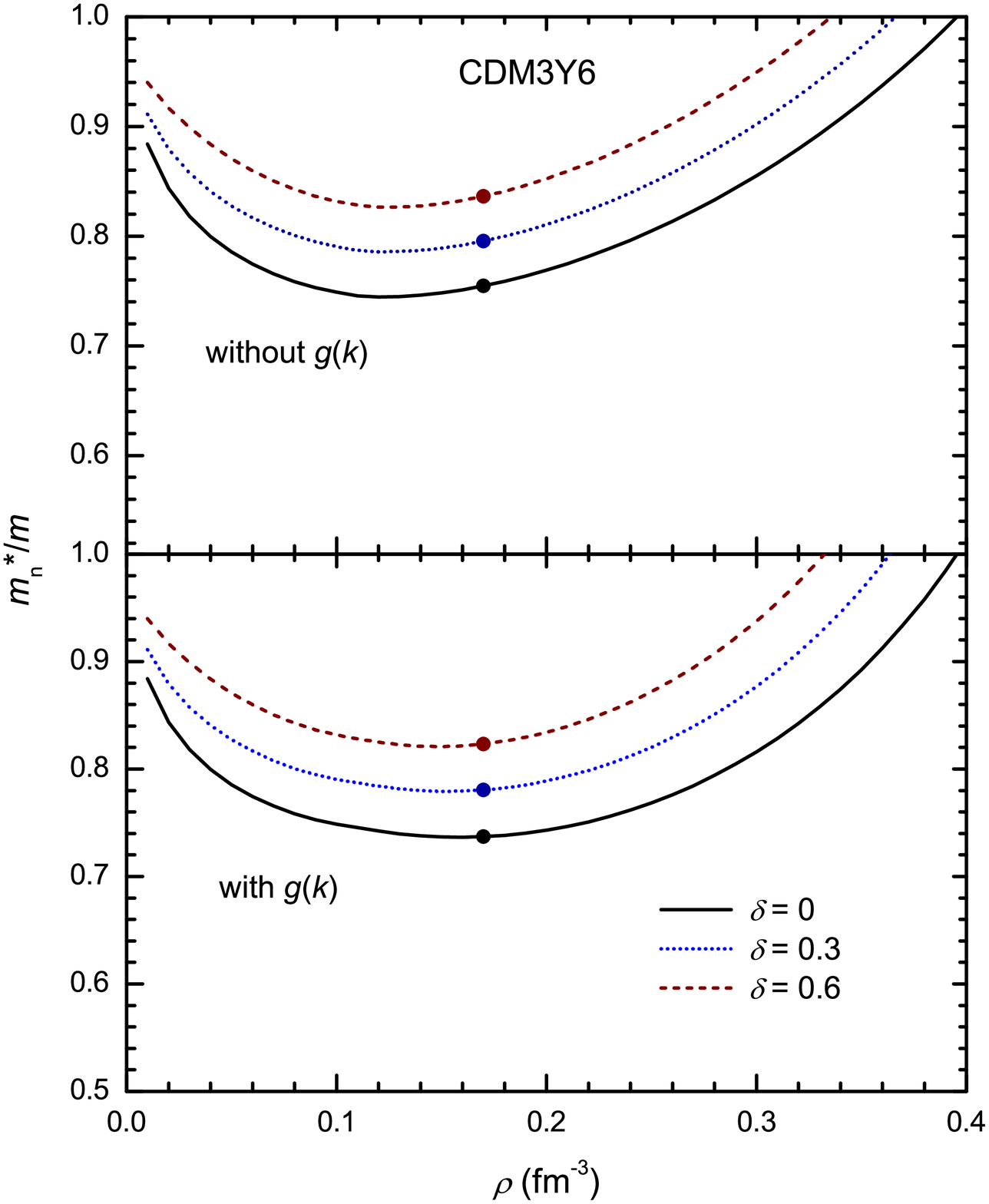}\vspace*{-0.5cm}
 \caption{(Color online) The same (HF+RT) results for the neutron effective mass
 as in the lower panel of Fig.~\ref{fEffm}, obtained with (lower panel) or without 
(upper panel) the modification of the high-momentum part of the s/p potential by the 
$g(k)$ function implied by the observed energy dependence of the nucleon OP.} \label{fmk}
\end{figure}
\begin{figure}[bht] \vspace*{-0.5cm}
\includegraphics[width=0.8\textwidth]{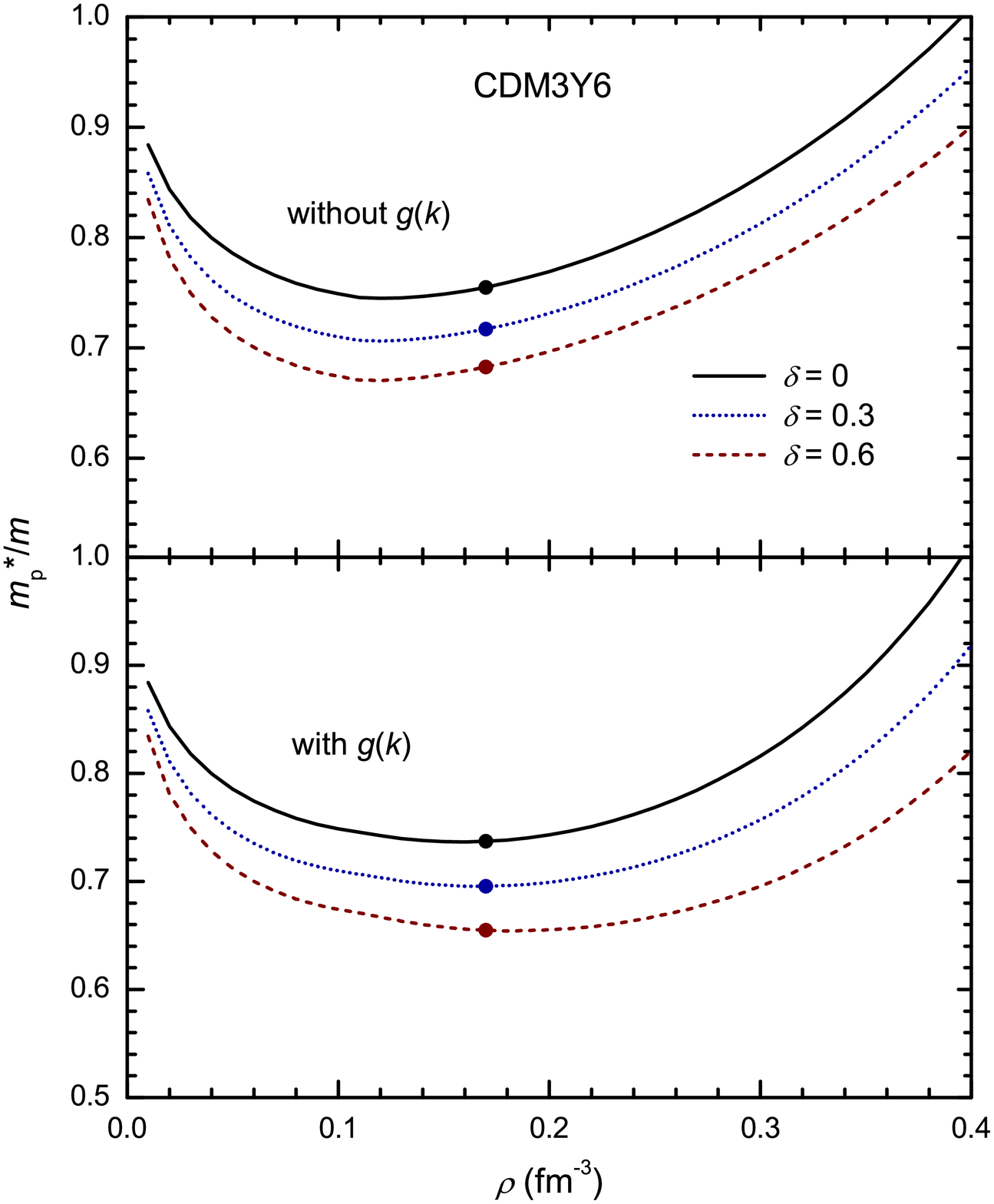}\vspace*{-0.5cm}
 \caption{The same as Fig.~\ref{fmk} but for the proton effective mass.} \label{fmk1}
\end{figure}
In the present work we do not intend to explore this issue over such a wide scope, 
but focus briefly on the effect of the RT in the determination of the nucleon 
effective mass within the extended HF formalism as well as the effect caused by
the modification of the high-momentum part of the s/p potential to match the  
observed energy dependence of the nucleon OP. The obtained neutron and proton 
effective masses are shown in Figs.~\ref{fEffm} and Figs.~\ref{fEffm1}, respectively,
and one can see that the RT enhances $m^*_\tau$ substantially at high NM densities. 
A similar behavior of the nucleon effective mass can also be seen in the results 
of a recent microscopic BHF calculation by Baldo {\it et al.} \cite{Bal14}, where 
the RT originating from three-body force drastically enhances the nucleon effective 
mass at the high NM densities (see, e.g., Fig.~1 of Ref.~\cite{Bal14}). Like found 
in Sec.~\ref{sec2} for the nuclear symmetry energy, the modification of the 
high-momentum part of the (HF+RT) s/p potential by the $g(k)$ function implied 
by the observed energy dependence of the nucleon OP changes slightly the slope 
of the density dependence of the nucleon effective mass at the high NM densities 
(see Figs.~\ref{fmk} and \ref{fmk1}). Although the nucleon effective mass is still
poorly known at the high NM densities and/or the large n/p asymmetries, the empirical 
$m^*/m$ value in the symmetric NM is known to be about 0.73 \cite{Hs83} at the 
saturation density $\rho_0$. In a good agreement with that empirical data, our HF 
and HF+RT results obtained with the CDM3Y6 interaction (see Fig.~\ref{fmk}) 
give $m^*/m\approx 0.734$ and 0.755, respectively. The modification of the high-momentum 
part of the (HF+RT) single-nucleon potential by the $g(k)$ function reduces this
result slightly to $m^*/m\approx 0.737$ at $\rho=\rho_0$.       

\begin{figure}[bht] \vspace*{-0.5cm}
\includegraphics[width=0.8\textwidth]{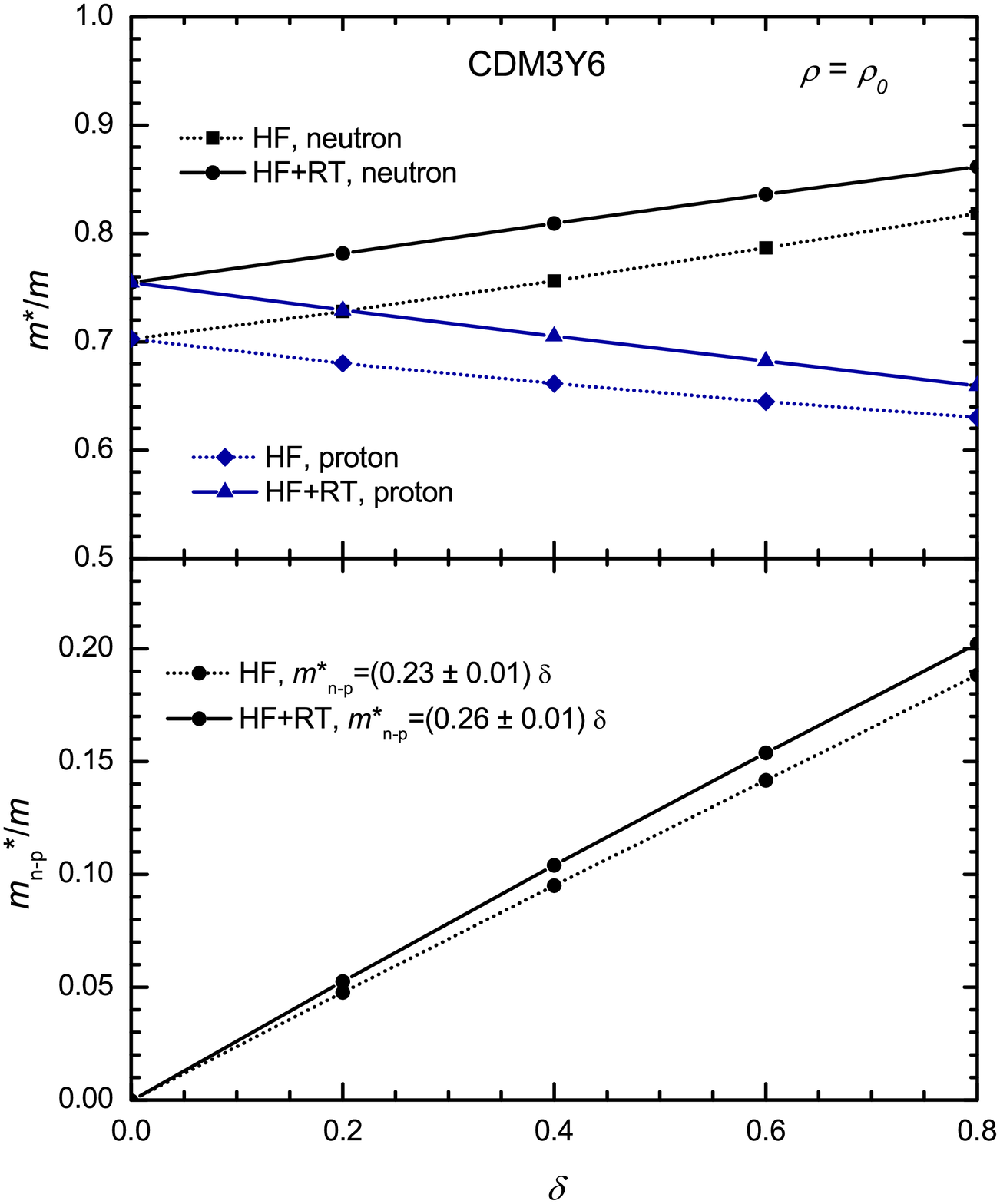}\vspace*{-0.5cm}
 \caption{(Color online) Neutron and proton effective masses (upper panel) 
and their splitting (lower panel) obtained at $\rho=\rho_0$ and the different n/p 
asymmetries $\delta$. Both the HF and HF+RT results show a well defined 
linear $\delta$-dependence of the n/p effective mass splitting (\ref{es5}).} 
 \label{fmd}
\end{figure}
\begin{figure}[bht] \vspace*{-0.5cm}
\includegraphics[width=0.8\textwidth]{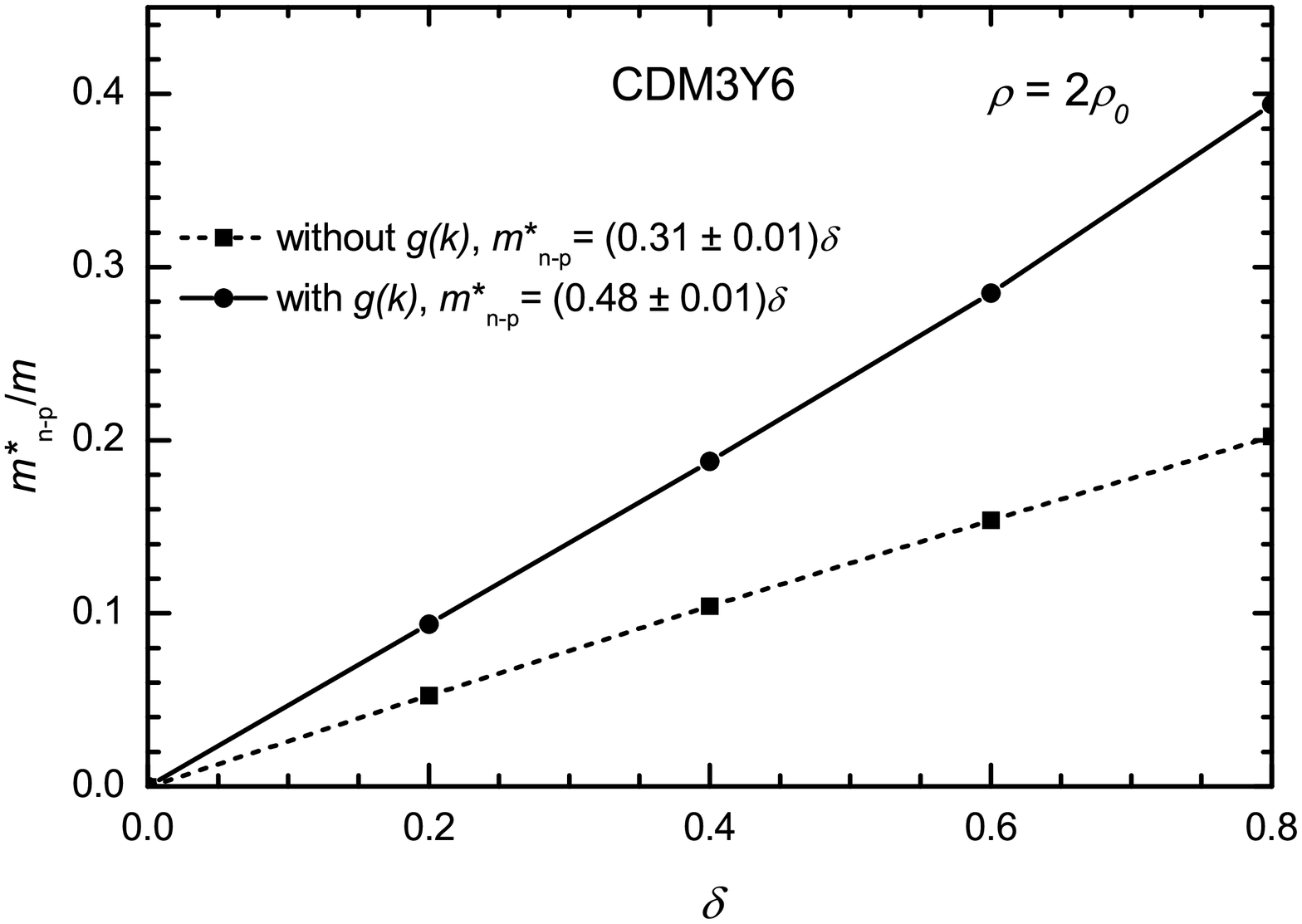}\vspace*{-1cm}
 \caption{$\delta$-dependence of the neutron-proton effective 
 mass splitting obtained at $\rho=2\rho_0$. The HF+RT results were obtained with 
(solid line) or without (dashed line) the modification of the high-momentum part 
of the s/p potential by the $g(k)$ function determined from the observed energy 
dependence of the nucleon OP.} \label{fmd2}
\end{figure}
With quite a good agreement of the calculated nucleon effective mass with the
empirical data at $\delta=0$, it is of interest to consider further the 
$\delta$-dependence of the $m^*/m$ value. The HF results obtained at the saturation 
density $\rho_0$ are shown in Fig.~\ref{fmd}. Given the equal RT contribution to 
each of the neutron- and proton s/p potential, the derivative with respect to the 
nuc;eon momentum gives about the same RT contribution of around 10\% to the 
neutron- and proton effective masses (upper panel of Fig.~\ref{fmd}), and the 
n/p effective mass splitting (lower panel of Fig.~\ref{fmd}) is rather weakly
affected by the rearrangement effect. On can see that both the standard HF and 
extended HF+RT results show a well defined linear $\delta$-dependence of the 
n/p effective mass splitting (\ref{es5}). Our final HF+RT results give 
$m^*_{n-p}(\rho_0,\delta)\approx (0.26\pm 0.01)\delta$ which is well inside 
the empirical boundary $m^*_{n-p}(\rho_0,\delta)\approx (0.27\pm 0.35)\delta$, 
established from the analyses of the terrestrial nuclear physics experiments 
and astrophysical observations \cite{Li13}. Our result is, however, lower than that 
estimated recently, $m^*_{n-p}(\rho_0,\delta)\approx (0.41\pm 0.15)\delta$, from 
the phenomenological (isospin-dependent) nucleon OP \cite{Li15} determined from 
the extensive optical model analysis of a large data set of the nucleon elastic 
scattering. The uncertainty of $0.01\delta$ in the HF+RT result is \emph{not} 
statistical error associated with the uncertainties of the model ingredients, 
but the uncertainty in adopting a linear $\delta$- dependence of the n/p effective
mass splitting (\ref{es5}) in Fig.~\ref{fmd}. This fact indicates simply that at 
$\delta <1$, when the parabolic approximation (\ref{es1}) and (\ref{es3}) is 
reasonable for the nuclear symmetry energy, the first-order symmetry term in 
the expansion of the single-nucleon potential over $\delta$ (see, e.g., 
Ref.~\cite{Che12}) contributes overwhelmingly to the determination 
of the $m^*_{n-p}(\rho,\delta)$ value. 
  
With larger Fermi momentum $k_F$ at the high NM densities, the modification of the 
momentum dependence of the s/p potentials by the $g(k_F)$ function determined
from the observed energy dependence of the nucleon OP should be taken into account 
in the HF+RT calculation of the n/p effective mass splitting. 
One can see in Fig.~\ref{fmd2} that $m^*_{n-p}(\rho,\delta)$ becomes larger 
at $\rho=2\rho_0$, and is enhanced further by more than 50\% when the modification 
of the high-momentum part of the s/p potential is taken into account. The behavior 
of the nucleon effective mass (\ref{es4}) and the n/p effective mass splitting 
(\ref{es5}) at the supranuclear densities is still poorly known, and the difference 
in the HF+RT results for the n/p effective mass splitting, 
$m^*_{n-p}(\rho,\delta)\approx (0.48\pm 0.01)\delta$ at $\rho=2\rho_0$ compared 
to $(0.26\pm 0.01)\delta$ at $\rho=\rho_0$, is quite significant and should be 
of interest for the nuclear astrophysical studies. 

\section{Folding model of the nucleon optical potential}
Given the substantial rearrangement effects to the nucleon OP found above 
in the extended HF calculation of the NM, it is of high interest to study 
these effects in the many-body calculation of the nucleon OP of the finite nuclei.
As such, the folding model (more precisely the single-folding model) has been 
proven to be a very effective tool to estimate the nucleon OP \cite{Kho02,Bri78,Amos}. 
It can be seen from the basic folding formulas that this model generates 
the first-order term of the microscopic OP defined in Feshbach's formalism 
of nuclear reactions \cite{Fe92}, based on the nucleon degrees of freedom. 
The success of the single-folding approach in the description of the elastic
\nA scattering data measured for the targets in the different mass regions 
suggests that the first-order term of the Feshbach's microscopic OP is indeed the 
dominant part of the nucleon OP. 

In the single-folding approach, the central OP of the $\tau$-kind nucleon 
incident on the target $A$ at the energy $E$ is evaluated as a HF-type 
potential \cite{Kho02}, using an appropriate, energy- and density dependent 
effective NN interaction $v_{\rm c}(\rho,E)$
\begin{eqnarray}
  U^{\rm fold}_{\tau}(E,R)&=&\sum_{j\in A}[\langle \bm{k}_\tau,j
	|v^{\rm D}_{\rm c}(\rho,E)|\bm{k}_\tau,j\rangle +\langle \bm{k}_\tau,j
	|v^{\rm EX}_{\rm c}(\rho,E)|j,\bm{k}_\tau\rangle], \nonumber \\
	&=& \sum_{j\in A}\langle \bm{k}_\tau,j
	|v_{\rm c}(\rho,E)|\bm{k}_\tau,j\rangle_\mathcal{A}, \label{ef1}
\end{eqnarray}
where $|j\rangle$ is the single-nucleon wave function of the $j-$th nucleon 
of the target. The antisymmetrization $\mathcal{A}$ of the \nA system is done 
by taking into account explicitly the knock-on exchange effects. As a result, 
the exchange term of $U_{\tau}$ becomes nonlocal in the coordinate 
space \cite{Amos}. An accurate local approximation is usually made by 
treating the relative motion locally as a plane wave \cite{Kho02,Bri78}, 
and the \emph{local} energy dependent folded potential (\ref{ef1}) is obtained
as an explicit function of the \nA distance $R$ and local momentum $k(R)$ 
of the incident nucleon. 

At the low incident energies, the pair-wise interaction between the incident 
nucleon and the nucleons bound in the target can induce certain rearrangement 
of the s/p configurations of the target nucleons that can be observed 
experimentally in the nucleon removal reactions \cite{Hs75}. In terms of the 
\nA interaction, such a rearrangement effect is expected to affect also the shape 
and strength of the \nA OP (\ref{ef1}), constructed in the folding model on the 
HF level from the s/p wave functions of the target nucleons. On the other 
hand, if we make a \emph{local density approximation} (LDA) for the s/p potential 
(\ref{Uhft}) by replacing the plane waves $|\bm{k}'\sigma'\tau'\rangle$ by the s/p 
wave functions $|j\rangle$ of the target nucleons, then the resulted potential 
is just the HF-type folded potential (\ref{ef1}) added by a rearrangement 
term, through the contribution of the $\Delta v_{\rm c}$ term. Thus, the total 
(central) nucleon OP now becomes   
\begin{equation}
  U_{\tau}(E,R)=U^{\rm HF}_{\tau}(k,R)+U^{\rm RT}_{\tau}(k,R)=
	\sum_{j\in A}\langle \bm{k}_\tau,j|v_{\rm c}(\rho,E)+\Delta v_{\rm c}(\rho)|
	\bm{k}_\tau,j\rangle_\mathcal{A}, \label{ef2}
\end{equation}
with $U^{\rm RT}$ originated from the explicit density dependence of the 
CDM3Yn interaction (\ref{CDM3Y}). In the standard HF formalism for the nucleon
mean-field potential, the HF potential for an unbound (scattering) nucleon has
been assumed years ago \cite{Vin76,Ber79} as the first-order term of the microscopic
OP for the low-energy elastic nucleon scattering, in about the same scheme as that
of the Feshbach's microscopic OP \cite{Fe92}. Applying the variational HF method to 
obtain the scattering equation for the \nA system, a rearrangement contribution 
to the HF potential appears naturally if the effective NN interaction is density
dependent \cite{Nak06}. Given the rearrangement contribution to the density dependence
of the CDM3Yn interaction determined above in the HF study of the NM, it is now 
possible to consistently include the RT into the folding calculation
of the \nA potential (\ref{ef1}). 

\begin{figure}[bht] \vspace*{-1.5cm}
\includegraphics[width=0.8\textwidth]{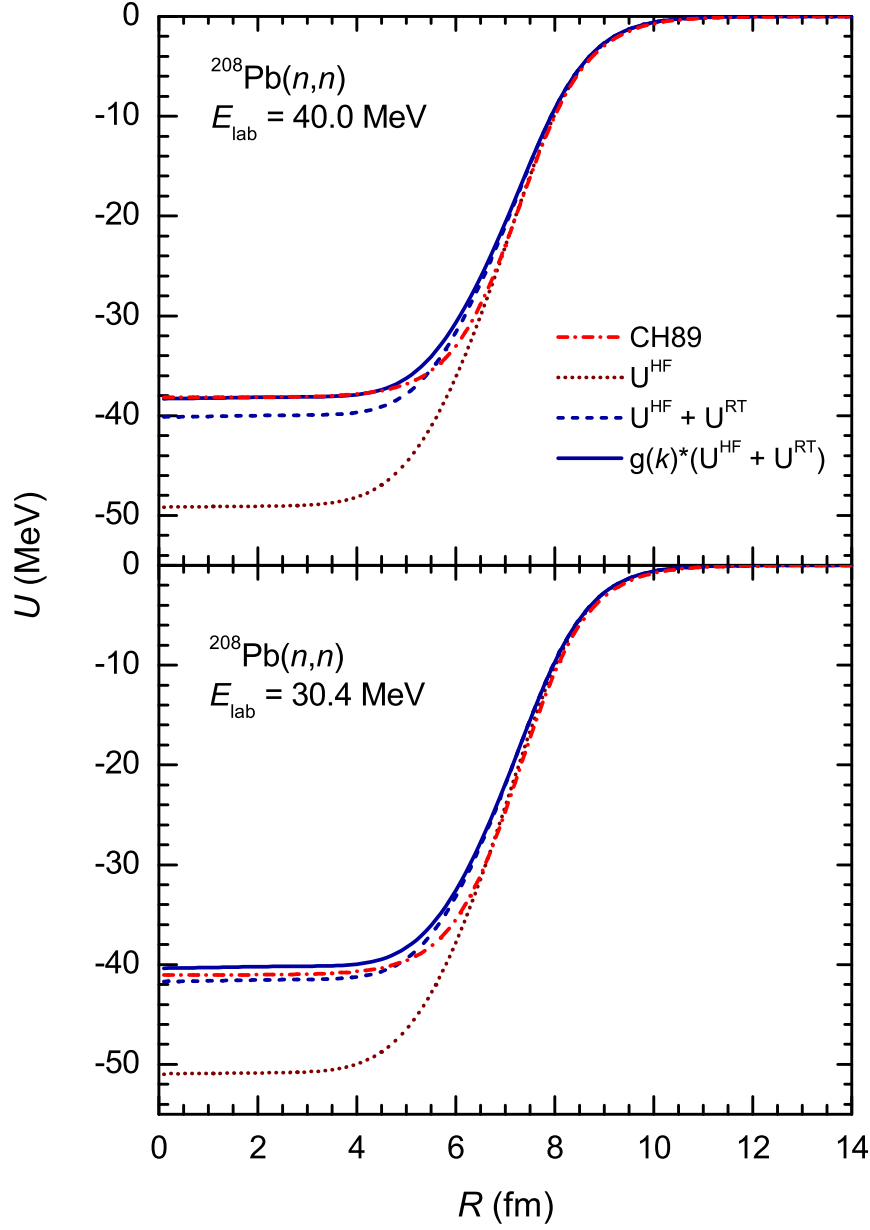}\vspace*{-1.5cm}
 \caption{(Color online) Real \nPb optical potential at the neutron incident
energies of 30.4 and 40 MeV predicted by the single-folding calculation 
(\ref{ef4})-(\ref{ef5}) using the CDM3Y6 interaction, with and without the 
contributions of the RT and the scaling $g\big(k(E,R)\big)$ function. CH89 
is the phenomenological OP taken from the global systematics 
by Varner {\it et al.} \cite{Va91}.} \label{fold1}
\end{figure}
Making explicit the neutron and proton s/p wave functions in Eq.~(\ref{ef2}) and
using the same isospin-dependent CDM3Yn interactions (\ref{CDM3Y}), we obtain 
the real nucleon OP explicitly in terms of the (energy dependent) isoscalar and 
isovector parts as 
\begin{equation}
 U_{\tau}(E,R)=U_{\rm IS}(E,R)\pm U_{\rm IV}(E,R), \label{ef3}
\end{equation}
where (-) sign pertains to the proton OP and (+) sign to the neutron OP. 
\begin{eqnarray}
 U_{\rm IS}(E,R) &=& g\big(k(E,R)\big)\int\Big[F_0\big(\rho(\bm{r})\big)+
  \Delta F_{0}\big(\rho(\bm{r})\big)\Big] 
\Big\{[\rho_n(\bm{r})+\rho_p(\bm{r})]v^{\rm D}_{00}(s) \nonumber\\
 & + & [\rho_n(\bm{R},\bm{r})+\rho_p(\bm{R},\bm{r})] 
 v^{\rm EX}_{00}(s)j_0\big(k(E,R)s\big)\Big\}d^3r, \label{ef4} \\
 U_{\rm IV}(E,R) &=& g\big(k(E,R)\big)\int\Big[F_1\big(\rho(\bm{r})\big)+
 \Delta F_1\big(\rho(\bm{r}),T_z(\bm{r})\big)\Big]  
\Big\{[\rho_n(\bm{r})-\rho_p(\bm{r})]v^{\rm D}_{01}(s) \nonumber\\
 & + & [\rho_n(\bm{R},\bm{r})-\rho_p(\bm{R},\bm{r})] 
v^{\rm EX}_{01}(s)j_0\big(k(E,R)s\big)\Big \}d^3r. \label{ef5} 
\end{eqnarray}
Here, $\rho_\tau(\bm{r},\bm{r}')$ is the nonlocal s/p density matrix of
the target with $\rho_\tau(\bm{r})\equiv\rho_\tau(\bm{r},\bm{r})$, 
and $k(E,R)$ is the local momentum of the incident nucleon determined 
self-consistently from the total real nucleon OP as     
\begin{equation}
 k(E,R)=\sqrt{\frac{2\mu}{\hbar^2}[E_{\rm c.m.}-U_{\tau}(E,R)]}. \label{ef6}
\end{equation} 
Similarly to the expression of the nucleon OP in the NM limit (\ref{Utotalg}), the
folded \nA potential (\ref{ef3}) also depends explicitly on the momentum (\ref{ef6}) 
of the incident nucleon through the (\emph{localized}) exchange term. 
To estimate properly the contribution of the rearrangement effects, the IS and IV 
density dependences of the RT in Eqs.~(\ref{ef4})-(\ref{ef5}) are determined in the 
LDA from the $\Delta F_0(\rho)$ and $\Delta F_1(\rho,\delta)$ values given by the 
HF study of the NM (\ref{dF}), accurately interpolated for the local nuclear 
density $\rho=\rho(\bm{r})=\rho_n(\bm{r})+\rho_p(\bm{r})$ and the neutron-proton 
asymmetry $\delta=T_z(\bm{r})=[\rho_n(\bm{r})-\rho_p(\bm{r})]/\rho(\bm{r}).$ 

\begin{figure}[bht] \vspace*{-0.5cm}
\includegraphics[width=1.0\textwidth]{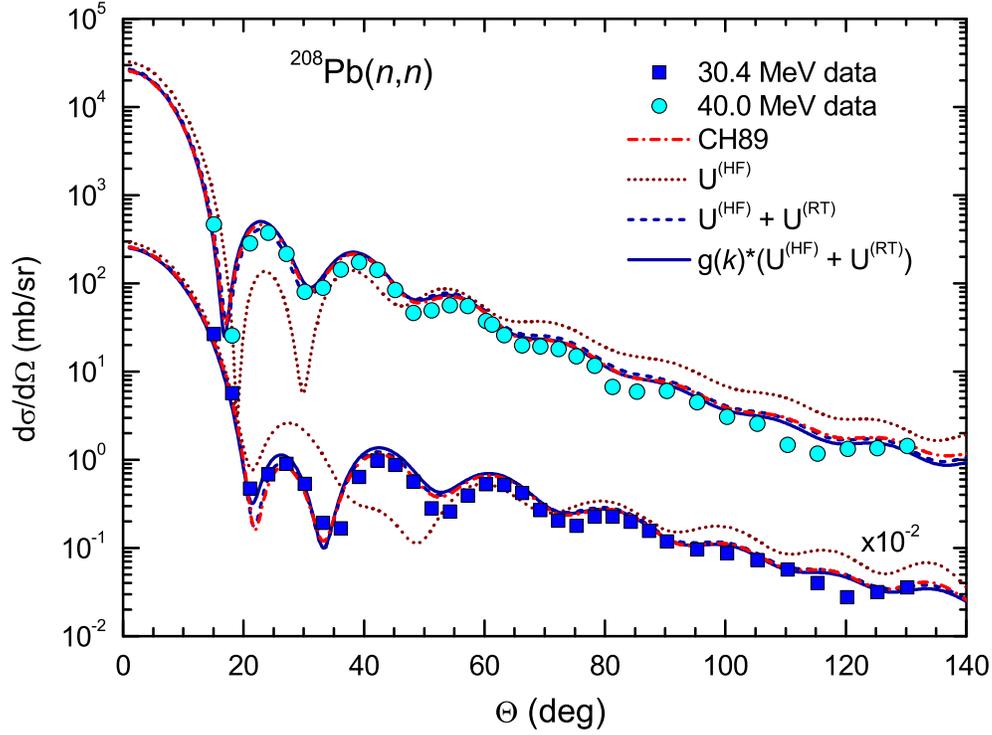}\vspace*{-0.5cm}
 \caption{(Color online) Elastic \nPb scattering cross sections at the neutron 
incident energies of 30.4 and 40 MeV obtained with the real OP's shown in 
Fig.~\ref{fold1}, in comparison with the data measured by DeVito {\it et al.} 
\cite{Dev12}. The imaginary and spin-orbital parts of the OP taken from the global 
systematics CH89 \cite{Va91} were used in the OM calculation.} \label{fold2}
\end{figure}
It can be seen from Eqs.~(\ref{ef4})-(\ref{ef5}) that the energy dependence of the 
real nucleon OP (\ref{ef3}) is determined entirely by the local momentum of the 
incident nucleon $k(E,R)$ appearing in the exchange potential as well as in 
the $g\big(k(E,R)\big)$ function. Given the $g(k)$ function determined above 
in the HF study of the NM based on the observed energy dependence of the nucleon 
OP, each local scaling factor $g\big(k(E,R)\big)$ of the folded potential is 
interpolated from the $g(k)$ function (see Fig.~\ref{fgk}) at the local 
momentum $k=k(E,R)$. As a result, $g\big(k(E,R)\big)$ can now be considered 
as the explicit energy (or momentum) dependence of the density dependent CDM3Yn 
interaction (\ref{CDM3Y}), locally consistent with the nucleon mean-field 
potential (\ref{ef3}). This is an essential improvement of the present formulation 
of the single-folding model, compared to the earlier applications of the folding 
model (see, e.g., Ref.~\cite{Kho02}) where a constant factor $g(E)\approx 1-0.0026E$ 
was used to scale the CDM3Y6 interaction. Because of the self-consistent 
determination of the $g\big(k(E,R)\big)$ function and contribution of the RT through 
$\Delta F_{0}\big(\rho(\bm{r})\big)$ and $\Delta F_1\big(\rho(\bm{r}),T_z(\bm{r})\big)$,
the single-folding calculation (\ref{ef4})-(\ref{ef5}) becomes more cumbersome and 
time consuming compared with the earlier version \cite{Kho02} of the folding model. 

\begin{figure}[bht] \vspace*{-1cm}
\includegraphics[width=0.8\textwidth]{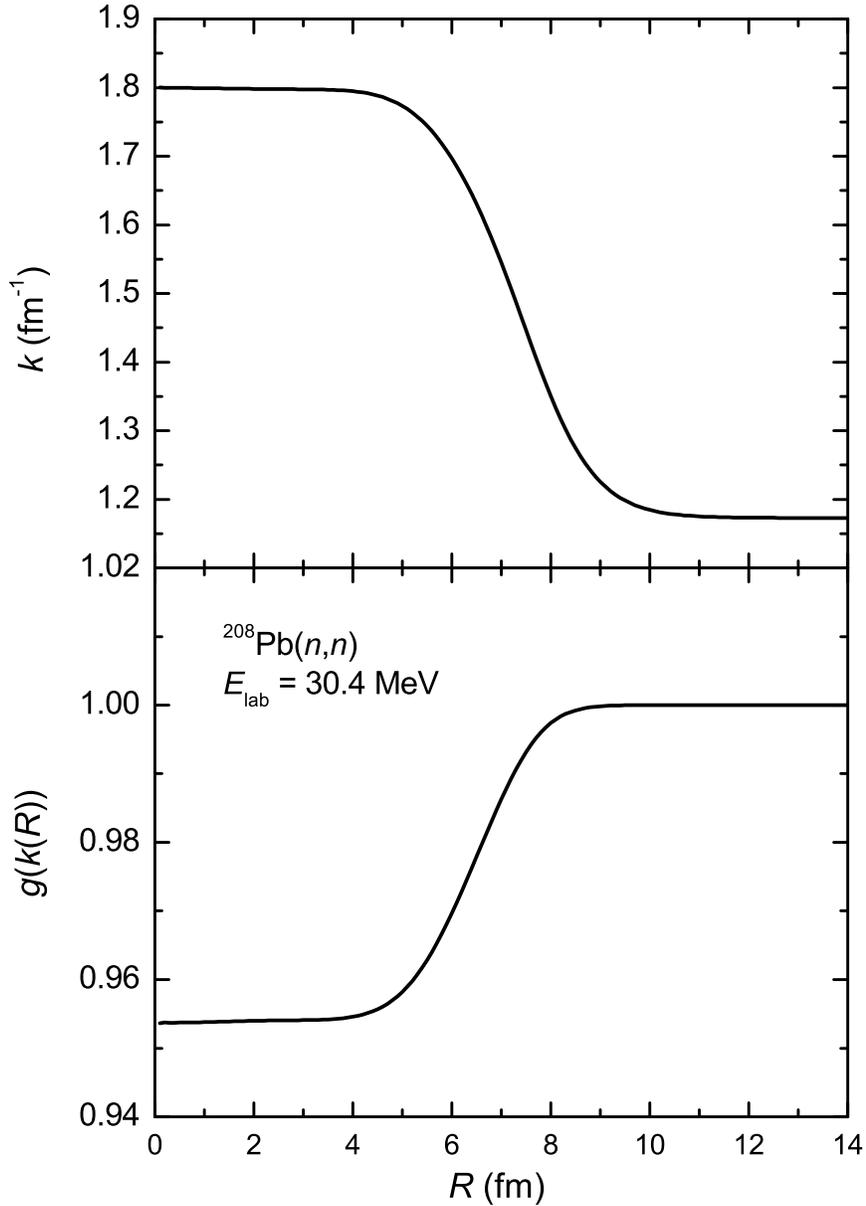}\vspace*{-1.5cm}
 \caption{Local momentum $k(E,R)$ of the incident neutron (upper panel) and 
the scaling function $g\big(k(E,R)\big)$ (lower panel) determined self-consistently 
from the HF+RT real folded \nPb potential obtained with the CDM3Y6 interaction at 
the neutron energy $E=30.4$ MeV.} \label{fold3}
\end{figure}
Although a comprehensive folding model study of the elastic \nA scattering based on
the new formulation of the model should be the subject of a separate study, we have
considered selectively in the present work the data of the elastic neutron scattering 
on the lead target, measured at the incident energies of 30.4 and 40 MeV \cite{Dev12}. 
Given no Coulomb interference and the energies close to the Fermi energy ($E_F\approx 
38.7$ MeV obtained with the Fermi momentum $k_{F}\approx 1.36$ fm$^{-1}$), the 
considered elastic neutron scattering data should be a good test ground for the
improved single-folding approach suggested in the present work, with the rearrangement 
effects consistently taken into account. For the $^{208}$Pb target we have used 
the empirical neutron and proton densities deduced from the high-precision elastic 
proton scattering at 800 MeV by Ray {\it et al.} \cite{Ray78,Ray78c}. These densities 
are available in the analytical form, and they have been used recently in our folding 
model analysis \cite{Loc14} of the charge-exchange $(^3$He,$t$) scattering to the isobar 
analog state of the target, to determine the thickness of the neutron skin in $^{208}$Pb.

From the folded \nPb potentials shown in Fig.~\ref{fold1} one can see that the 
standard folding method \cite{Kho02} gives a rather deep HF-type folded potential. 
After the RT is taken properly into account, the HF+RT folded potential becomes 
substantially shallower, much closer to the empirical Woods-Saxon potential given 
by the global CH89 parametrizations of the nucleon OP \cite{Va91}, which was proven to be 
accurate for the elastic nucleon scattering from medium and heavy targets at the incident 
energies below 100 MeV. The agreement of the HF+RT folded potential with the empirical CH89 
potential becomes better, especially at the neutron energy of 40 MeV, when the scaling 
$g\big(k(E,R)\big)$ function is consistently taken into account. To further test 
the folded \nPb potentials shown in Fig.~\ref{fold1}, they were used (without any further
renormalization of their strength) as the real OP in the standard optical model (OM) 
calculation of the elastic \nPb scattering at 30.4 and 40 MeV using the code ECIS06 
written by Raynal \cite{Ray06}, with the imaginary 
and spin-orbital parts of the OP taken from the global systematics CH89 \cite{Va91}. 
From the comparison of the calculated cross sections with the measured data \cite{Dev12} 
in Fig.~\ref{fold2} one can see that a very good OM description of the data has been 
obtained with the real folded OP after the RT is included. Although the OM fit is 
marginally improved after the HF+RT folded potential is scaled with the scaling 
function $g\big(k(E,R)\big)$, it is important to note that this scaling function
is resulted from the realistic momentum dependence of the s/p potential in the NM 
discussed in Sec.~\ref{sec1}. The local momentum $k(E,R)$ of the incident 
neutron is largest in the center, and approaches its asymptotic value at the potential
surface (determined for a free neutron whose kinetic energy is equal the incident 
energy) (see upper panel of Fig.~\ref{fold3}). As a result, the $g\big(k(E,R)\big)$ 
value is ranging smoothly from about 0.96 at small radii to unity at the potential 
surface, which is a rather mild mean-field effect on the shape of the folded 
\nA potential. In any case, such a scaling function is physically much more consistent 
compared to a constant $g(E)$ factor used in the earlier version of the folding 
model \cite{Kho02} at each nucleon incident energy. The standard HF folded potential 
is too deep in the center and cannot deliver a good OM description of the data 
(see Fig.\ref{fold2}) unless its strength is \emph{renormalized} by a factor 
$N_R\approx 0.85$. A renormalization factor $N_R<1$ of the real folded OP was 
often obtained in the earlier folding model analyses  of the elastic nucleon 
scattering \cite{Kho07}, and the lack of the contribution from the RT is likely 
the main reason. 

With a very good OM description of the considered elastic neutron scattering data
by the HF+RT folded potential, the reliability of the folding model in predicting 
the \nA OP seems much improved. In view of the new folding formalism suggested in
the present work, a systematic folding model analysis of the elastic and inelastic 
\nA scattering over a wide range of energies should be of high interest.                  

\section{Summary}
A consistent HF study of the asymmetric NM has been done using the CDM3Y3 
and CDM3Y6 density dependent versions of the M3Y-Paris interaction, with the
focus on the rearrangement term of the s/p potential that arises naturally
when the Hugenholtz-van Hove theorem is applied to calculate s/p energy from 
the total NM energy. Based on the exact expression of the RT of the 
isospin dependent s/p potential given by the HvH theorem at each NM density and 
the empirical energy dependence of the nucleon OP observed over a wide range 
of energies, a simple method has been proposed to account effectively for the 
momentum dependence of the RT of the s/p potential in the standard HF scheme, 
with an explicit contribution of the RT added to the density dependence 
of the CDM3Yn interaction (\ref{CDM3Y}). The HF+RT s/p potentials obtained at 
the different NM densities and neutron-proton asymmetries agree reasonably with 
those predicted by the microscopic BHF calculations of the NM that have the 
higher-order rearrangement term properly included \cite{Zuo99,Vi13}.  

Given a direct link between the s/p potential in the NM and the nuclear symmetry 
\cite{Xu14,Zuo14}, we have determined anew the parameters of the isovector densiy
dependence of the CDM3Yn interaction (\ref{CDM3Y}) by matching our HF+RT results
for the IV part of the nucleon OP in the NM with those of the BHF calculation by the 
JLM group \cite{Je77,Lej80}. Using these new parameters, the calculated nuclear 
symmetry energy $S(\rho)$ agrees nicely with the latest empirical constraints 
at $\rho\leq\rho_0$ as well as the results of the ab-initio calculations of the
asymmetric NM at $\rho > \rho_0$. With the high-momentum part of the s/p potential
modified based on the observed energy dependence of the nucleon OP, the nuclear
symmetry energy $S(\rho)$ obtained in the parabolic approximation from the 
difference between the neutron and proton s/p energies turns out to be in a 
very good agreement with that given by the ab-initio calculations over a wide 
range of the NM densities. This result indicates that one might indirectly learn 
about the density dependence of the nuclear symmetry energy from the extensive 
OM studies of the elastic \nA scattering at low, medium, and intermediate energies. 

The momentum dependence of the s/p potential obtained in the HF+RT calculation 
has a very well defined linear $\delta$-dependence of the n/p effective mass 
splitting (\ref{es5}), with $m^*_{n-p}(\rho_0,\delta)\approx (0.26\pm 0.01)\delta$ 
that agrees well with the empirical constraint from the recent analysis of the 
terrestrial nuclear physics experiments and astrophysical observations \cite{Li13}. 
The $m^*_{n-p}(\rho,\delta)$ value was found to be readily increased with the 
increasing NM density up to $2\rho_0$, while retaining its linear $\delta$-dependence. 
Although there is no empirical constraint for the n/p effective mass splitting 
at the high NM densities to compare with, we have shown here that a proper treatment 
of the RT and a realistic momentum dependence of the single-nucleon potential are 
prerequisites for the determination of $m^*_{n-p}(\rho,\delta)$ at the different 
NM densities that can be of interest for the nuclear astrophysical studies.   

A very important milestone of the present work is that the proper treatment 
of the rearrangement effects and momentum dependence of the nucleon mean-field 
potential in the HF calculation of the NM has lead us to the important
physics inputs that enable a consistent inclusion of the RT into the HF-type 
folding model calculation of the nucleon OP of the finite nuclei in the same 
mean-field manner. The contribution of the RT has been shown, in an application 
of the extended folding model to study the elastic \nPb scattering, 
to be vital in obtaining the realistic shape and strength of the real nucleon OP.
The predicting power of the folding model for the nucleon OP seems much improved. 
A systematic folding model analysis of the elastic and inelastic \nA scattering 
over a wide range of energies is now planned with the extended folding formalism.

\section*{Acknowledgments}
We thank Bao-An Li and Chang Xu for their helpful communications. The present 
research has been supported by the National Foundation for Scientific and Technological
Development (NAFOSTED Project No. 103.04-2014.76).

\end{document}